\documentclass[pra,twocolumn,showpacs,preprintnumbers,amsmath,amssymb,superscriptaddress]{revtex4-1}

\usepackage{amsfonts}
\usepackage{amsmath}
\usepackage{amssymb}
\usepackage{graphicx}
\usepackage[nonscript]{accents}
\usepackage{color}
\newcommand{\cnb}[1]{{\color{black} #1}}

\newcommand{\rtxt}[1]{{\color{black} #1}}
\newcommand{\gtxt}[1]{{\color{black} #1}}
\newcommand{\btxtb}[1]{{\color{black} #1}}

\begin{document}

\title{Classical and quantum-linearized descriptions of \\degenerate optomechanical parametric oscillators}

\author{Sebastian Pina-Otey}
\affiliation{Max-Planck-Institut f\"{u}r Quantenoptik, Hans-Kopfermann-strasse 1, 85748
Garching, Germany}

\author{Fernando Jim\'{e}nez}
\affiliation{Zentrum Mathematik der Technische Universit\"{a}t M\"{u}nchen,
Boltzmann-strasse 3, 85747 Garching, Germany}

\author{Peter Degenfeld-Schonburg}
\affiliation{Technische Universit{\"a}t M{\"u}nchen, Physik Department, James Franck Str., 85748 Garching, Germany}

\author{Carlos Navarrete-Benlloch}
\email{carlos.navarrete@mpq.mpg.de}
\affiliation{Max-Planck-Institut f\"{u}r Quantenoptik, Hans-Kopfermann-strasse 1, 85748
Garching, Germany}

\begin{abstract}
Recent advances in the development of modern quantum technologies have opened the possibility of studying the interplay between spontaneous parametric down-conversion and optomechanics, two of the most fundamental nonlinear optical processes. Apart from practical reasons, such scenario is very interesting from a fundamental point of view, because it allows exploring the optomechanical interaction in the presence of a strongly quantum-correlated field, the spontaneously down-converted mode. In this work we analyze such problem from two approximate but valuable perspectives: the classical limit and the limit of small quantum fluctuations. We show that, in the presence of optomechanical coupling, the well-known classical phase diagram of the optical problem gets modified by the appearance of new dynamical instabilities. As for the quantum-mechanical description, we prove the ability of the squeezed down-converted field to cool down the mechanical motion not only to thermal but also to squeezed thermal mechanical states, and in a way that can be much less sensitive to parameters (e.g., detuning of the driving laser) than standard sideband cooling.
\end{abstract}


\maketitle

\section{Introduction}

Spontaneous parametric down-conversion is a process which occurs in crystals with second-order optical nonlinearity, where light at some frequency
$2\omega_{0}$ can be transformed into light at frequencies $\omega_{\mathrm{s}}$
and $\omega_{\mathrm{i}}$ such that $\omega_{\mathrm{s}}+\omega_{\mathrm{i}%
}\approx2\omega_{0}$ \cite{Boyd,BlueBook,NavarretePhDthesis,CarmichaelBook2}.
When the crystal is introduced in an optical cavity, what has the effect of
enhancing the nonlinear interaction and filter the fields all at once, we
obtain a so-called optical parametric oscillator (OPO), in which the down-converted
field starts oscillating in the cavity only once the power of the pumping
laser exceeds some threshold value (such that the nonlinear gain can
compensate for the cavity losses) \cite{Boyd,
NavarretePhDthesis,CarmichaelBook2}. These devices have found many
applications both in classical and quantum optics. In the classical case, they
are among the most tunable sources of light, allowing to transform laser light
into almost any (optical) frequency \cite{Boyd}. From a quantum point of view,
the down-converted photons show strong quantum correlations; particularly
relevant to this work is the degenerate optical parametric oscillator (DOPO),
in which down-converted photons have the same frequency $\omega_{\mathrm{s}%
}=\omega_{\mathrm{i}}\approx\omega_{0}$, and the corresponding output field
shows nearly-perfect quadrature squeezing when working close to threshold
\cite{BlueBook,NavarretePhDthesis,CarmichaelBook2}. Indeed, DOPOs are nowadays
the sources of the highest-quality squeezed light
\cite{Takeno07,Vahlbruch08,Eberle10,Mehmet10}, which can be used to increase
the sensitivity of measurements beyond the standard quantum limit
\cite{Ligo11,Goda08, Vahlbruch05, Treps03, Treps02}, or also to generate entangled
beams for quantum information purposes \cite{Braunstein05,
Weedbrook12,NavarreteQI}.

On the other hand, we have optomechanical systems, where some mechanical
degree of freedom is coupled to a light field via, e.g., radiation pressure
\cite{Kippenberg07,Marquardt09,Meystre13,Aspelmeyer13}. When the interaction
happens inside a cavity, the Lorentzian density of modes provided by the
resonator, together with the injection of a coherent laser field with the
proper detuning with respect to the cavity resonance, allows to cool down the
mechanical degree of freedom to its quantum mechanical ground state through
sideband cooling
\cite{Gigan06,Arcizet06,Schliesser06,Corbitt07,Thompson08,Wilson09,Teufel11,Chan11,Karuza12,Harris15,WilsonRae07,Marquardt07,Genes08}%
. Since the mechanical degree of freedom is usually a mesoscopic system formed
by many atoms, optomechanics provides a very promising platform where studying
the transition from the microscopic quantum world to our natural macroscopic
classical one, allowing, for example, to put bounds on collapse models
\cite{Marshall03,Kleckner08,Romero-Isart11,Romero-Isart12}. From a practical
point of view, apart from offering a new platform where performing traditional
quantum optical tasks such as the generation of squeezed light
\cite{Fabre94,Mancini94,Brooks12,Safavi13}, transparency windows
\cite{Weis10,Safavi11omit,Teufel11omit,Massel12,Karuza13}, or photon blockade
effects \cite{Rabl11}, optomechanical systems might be a perfect interface
between optical and microwave technologies, since mechanical degrees of
freedom couple to both electromagnetic scales
\cite{Stannigel10,Safavi11,Regal11,Taylor11,Wang12PRL,Wang12,Barzanjeh12,Bochmann13,Bagci14,Andrews14}.

As for actual optomechanical implementations, they come out in many different
forms \cite{Aspelmeyer13}: cavities with mirrors attached to cantilevers
\cite{Gigan06,Arcizet06} or in suspension \cite{Corbitt07}, flexible membranes
placed inside optical cavities \cite{Thompson08,Wilson09,Karuza12}, or
localized mechanical modes in photonic crystal cavities \cite{Chan11}, are
some examples. For our current purposes, the most relevant implementations
consist in (i) a whispery gallery mode resonator (a microtoroid or microdisk, for
example) where light circulates around its edge via total internal reflection, pushing the whole structure, hence exciting some of its mechanical modes \cite{Schliesser06,SchliesserPhDthesis}, and (ii) flexible drum-shaped capacitors coupled to superconducting LC resonators \cite{Teufel11,Teufel11omit,Palomaki13}.

From a fundamental point of view, the interplay between optomechanics and
down-conversion seems to be a natural and interesting problem to
study within the nonlinear quantum optics community. Recent theoretical results for the case in which the down-converted mode is seeded with an external field  predict that such \textit{stimulated} down-conversion process is able to enhance optomechanical cooling \cite{Agarwal09}, normal mode splitting \cite{Agarwal09bis}, and the sensitivity of mechanical quadrature measurements \cite{Marquardts15}, as well as generate mechanical squeezed states \cite{Agarwal16} or even bring the optomechanical interaction to the (single-quanta) strong coupling regime \cite{Nori15}. In a second-harmonic generation configuration (in which only the down-converted frequency is driven), multipartite optomechanical entanglement has also been predicted \cite{Xuereb12}.

Modern platforms capable of combining down-conversion and optomechanics in the same device have turned the motivation for studying such a scenario into a practical one. In particular, miniaturized whispering gallery mode resonators can be fabricated directly with the typical crystalline materials possessing second-order optical nonlinearity, such that light can be down-converted while circulating on the resonator \cite{Ilchenko03,Ilchenko04,Savchenkov07,Furst10,Furst10b,Hofer10,Furst11,Beckmann11,Werner12,Fortsch13,Marquardt13,Fortsch14,Fortsch14b}. A completely different, but equally realistic implementation could consist on a superconducting circuit in the degenerate parametric oscillation configuration \cite{Leghtas15} coupled to a drum-shaped capacitor acting as a the mechanical degree of freedom \cite{Teufel11,Teufel11omit,Palomaki13} (see also \cite{Nation15} for a recent circuit QED proposal). \btxtb{A third and natural option would consist in using a standard OPO cavity built with a movable micromirror similar to those in \cite{Gigan06,Arcizet06}.}

\begin{figure}[t]
\includegraphics[width=\columnwidth]{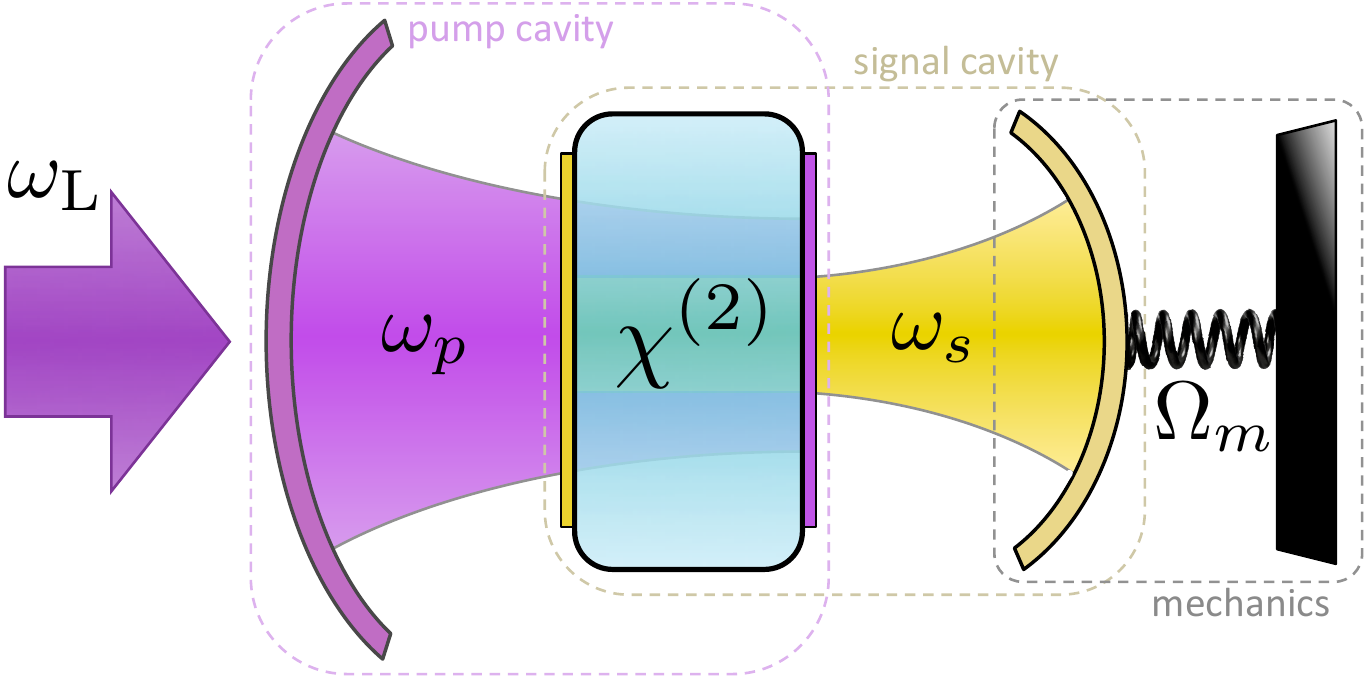} \caption{Sketch of the degenerate optomechanical parametric oscillator considered in this work. It
consists of a cavity containing a mode at frequency $\omega_\mathrm{p}$ (\textit{pump}) driven by a laser at frequency $\omega_\mathrm{L}$. The pump cavity shares a $\chi^{(2)}$ nonlinear crystal with another cavity containing a resonance at frequency $\omega_\mathrm{s}$ close to $\omega_\mathrm{L}/2$ (\textit{signal}), which can be then populated via degenerate down-conversion of the pumping laser in the crystal. One of the mirrors of the signal cavity is not fixed and can act then as a mechanical oscillator forced by the radiation pressure exerted from the light contained in the cavity. \btxtb{OPOs based on dual semi-monolithic cavities \cite{Gigan07} can implement this exact setup, but other platforms such as crystalline whispering gallery mode resonators \cite{Ilchenko03,Ilchenko04,Savchenkov07,Furst10,Furst10b,Hofer10,Furst11,Beckmann11,Werner12,Fortsch13,Marquardt13,Fortsch14,Fortsch14b} or superconducting circuits \cite{Leghtas15,Teufel11,Teufel11omit,Palomaki13,Nation15} are already at a point where the model analyzed in our work can be studied experimentally.}}
\label{Scheme}
\end{figure}

One of the most relevant questions in these scenarios concerns the effect that the spontaneously down-converted light, which shows strong quantum correlations, will have on the mechanical state. As a step towards understanding this question, in this work we analyze the system from a classical perspective, providing also a small glance at the quantum properties predicted by a linearization of quantum fluctuations around the classical solution. For simplicity, we stick to the degenerate case and assume that only the down-converted field is coupled to the mechanical mode. \btxtb{The latter is a natural situation in circuit setups (where the pump and down-converted modes are provided by different linear circuits) or when the movable mirror is introduced in an OPO that makes use of dual semi-monolithic designs \cite{Gigan07} (see Fig. \ref{Scheme}) to create independent cavities for the pump and down-converted fields; in the case of crystalline whispering gallery mode resonators it might require choosing a proper mechanical mode weakly coupled to the optical modes around the pump frequency.} From a fundamental point of view, this configuration provides probably the most exciting scenario, since the pump stays near-coherent for most of the DOPO parameters, and its coupling to the mechanics could mask the effects generated by the quantum-correlated down-converted field.

Concerning the classical limit, we show that, with respect the usual DOPO scenario \cite{Lugiato88,Pettiaux89,Fabre90}, the most relevant effect that the optomechanical interaction has is the generation of new dynamical instabilities, as well as the modification of the region of intensity bistability. On the other hand, regarding the quantum linearized theory, we first argue how it completely fails to capture the physics below threshold, where the optomechanical interaction is purely nonlinear and requires more sophisticated techniques to describe it, which is precisely what we did in a recent work \cite{Degenfeld15b}. Nevertheless, the linearized description provides reasonable predictions above threshold, and we use it to show that the down-converted field can have a deep impact on the mechanical state, but very different from that of the standard coherently-pumped optomechanical cavity. In particular, we show that it can cool down the mechanical motion in a way less sensitive to parameters (particularly detuning) than the traditional sideband cooling, even generating squeezed thermal mechanical states as the optomechanical interaction is increased. \btxtb{Let us remark that an understanding of the classical phase diagram is instrumental prior to performing a more accurate quantum analysis, as we already emphasized in Ref. \cite{Degenfeld15b}.}

The article is organized as follows. First, we introduce the model of the system,
which we have called \textit{degenerate optomechanical parametric oscillator} (DOMPO)
and we describe through a set of quantum Langevin equations. Next we study the
classical limit of the model, finding its steady states and analyzing their
stability. Finally, we apply the standard linearization technique to study the
quantum properties of the mechanical mode in regions of the parameter space
where the classical stationary solution is stable.

\section{The model}

Even though the actual implementation can differ from the simple picture
sketched in Fig. \ref{Scheme}, a DOMPO can be schematically understood as depicted there.
Consider a cavity pumped by a laser at frequency $\omega_{\mathrm{L}}$ close
to one of its resonances $\omega_{\mathrm{p}}$. This cavity (denoted by
\textit{pump cavity}) shares a second-order nonlinear crystal with another
cavity (\textit{signal cavity}) in which a resonance $\omega_{\mathrm{s}}$
close to $\omega_{\mathrm{L}}/2$ exists, and therefore can be populated via
down-conversion of pump photons in the crystal. Finally, one mirror of the
signal cavity can oscillate at frequency $\Omega_{\mathrm{m}}$, and acts then
as a mechanical oscillator which is forced via the radiation pressure exerted by
the signal field.

Let us denote by $\hat{x}_{\mathrm{m}}=\hat{Q}/Q_{0}$ the displacement of the
mirror ($\hat{Q}$) normalized to the zero-point position fluctuations
$Q_{0}=\sqrt{\hbar/2\Omega_{\mathrm{m}}M}$, where $M$ is the mirror's mass. We
also normalize the corresponding momentum $\hat{P}$ to the zero-point momentum
fluctuations $P_{0}=\hbar/Q_{0}$, obtaining the normalized momentum $\hat
{p}_{\mathrm{m}}=\hat{P}/P_{0}$. We define annihilation operators $\hat{a}%
_{j}$ for pump ($j=\mathrm{p}$) and signal ($j=\mathrm{s}$) photons, with
corresponding creation operators $\hat{a}_{j}^{\dag}$. These operators satisfy
the commutation relations $[\hat{x}_{\mathrm{m}},\hat{p}_{\mathrm{m}%
}]=2\mathrm{i}$, and $[\hat{a}_{j},\hat{a}_{l}^{\dagger}]=\delta_{jl}$, any
other commutator between them being zero. In a picture rotating at the laser
frequency $\omega_{\mathrm{L}}$ for the pump and $\omega_{\mathrm{L}}/2$ for
the signal, the physical processes described in the previous paragraph are
then captured by the Hamiltonian $\hat{H}=\hat{H}_{\mathrm{O}%
}+\hat{H}_{\mathrm{M}}+\hat{H}_{\mathrm{DC}}+\hat
{H}_{\mathrm{OM}}$, with \cite{BlueBook, NavarretePhDthesis,CarmichaelBook2,Aspelmeyer13}
\begin{subequations}
\begin{eqnarray}
\hat{H}_\mathrm{O}&=&-\sum_{j=\mathrm{p,s}}\hbar\Delta_{j}\hat{a}_{j}^{\dagger}\hat{a}_{j}+\mathrm{i}\hbar\mathcal{E}_{\mathrm{L}}(\hat{a}_{\mathrm{p}}^{\dagger}-\hat{a}_{\mathrm{p}}),
\\
\hat{H}_\mathrm{M}&=&\frac{\hbar\Omega_{\mathrm{m}}}{4}(\hat{x}%
_{\mathrm{m}}^{2}+\hat{p}_{\mathrm{m}}^{2}),
\\
\hat{H}_\mathrm{DC}&=&\mathrm{i}\hbar\chi(\hat{a}_{\mathrm{p}}\hat{a}_{\mathrm{s}}^{\dagger2}-\hat{a}_{\mathrm{p}}^{\dagger}\hat{a}_{\mathrm{s}}^{2}),
\\
\hat{H}_\mathrm{OM}&=&-g_{\mathrm{s}}\hat{a}_{\mathrm{s}}^{\dagger}\hat{a}_{\mathrm{s}}\hat{x}_{\mathrm{m}},
\end{eqnarray}
\end{subequations}
where $\Delta_{\mathrm{p}}=\omega_{\mathrm{L}}-\omega_{\mathrm{p}}$ and
$\Delta_{\mathrm{s}}=\omega_{\mathrm{L}}/2-\omega_{\mathrm{s}}$ denote the
detuning of the laser with respect to the pump and signal modes,
$\mathcal{E}_{\mathrm{L}}$ is the pump cavity's driving rate (proportional to the square root of the power of the external laser),
$\chi$ is the down-conversion rate (proportional to the crystal's nonlinear
susceptibility), and $g_{\mathrm{s}}$ is the opto-mechanical scattering rate
(which depends strongly on the particular implementation).

In addition to these coherent processes, the system is subject to incoherent
processes. In particular, we need to take into account the loss of
photons through the partially transmitting mirrors (open cavities), and the
coupling of the mechanical oscillator to its thermal environment, reaching some equilibrium
temperature $T$ in the absence of light. We choose to include these processes at the level of the
Heisenberg equations of motion, leading to the widely-used quantum Langevin
equations \cite{ZollerBook,WMbook}
%
\begin{eqnarray}
\frac{d\hat{x}_{\mathrm{m}}}{dt}  &=& \Omega_{\mathrm{m}}\hat{p}_{\mathrm{m}},
\\
\frac{d\hat{p}_{\mathrm{m}}}{dt}  &=& -\gamma_{\mathrm{m}}\hat{p}_{\mathrm{m}}-\Omega_{\mathrm{m}}\hat{x}_{\mathrm{m}}+2g_{\mathrm{s}}\hat{a}_{\mathrm{s}%
}^{\dagger}\hat{a}_{\mathrm{s}}+\sqrt{4\gamma_{\mathrm{m}}\bar{n}_{\mathrm{th}}}\hat{p}_{\mathrm{m,in}}(t), \nonumber
\\
\frac{d\hat{a}_{\mathrm{p}}}{dt}  &=& \mathcal{E}_{\mathrm{L}}-(\gamma
_{\mathrm{p}}-\mathrm{i}\Delta_{\mathrm{p}})\hat{a}_{\mathrm{p}}-\frac{\chi
}{2}\hat{a}_{\mathrm{s}}^{2}+\sqrt{2\gamma_{\mathrm{p}}}\hat{a}_{\mathrm{p}%
,\mathrm{in}}(t), \nonumber
\\
\frac{d\hat{a}_{\mathrm{s}}}{dt}  &=& -(\gamma_{\mathrm{s}}-\mathrm{i}%
\Delta_{\mathrm{s}}-\mathrm{i}g_{\mathrm{s}}\hat{x}_{\mathrm{m}})\hat
{a}_{\mathrm{s}}+\chi\hat{a}_{\mathrm{p}}\hat{a}_{\mathrm{s}}^{\dag}%
+\sqrt{2\gamma_{\mathrm{s}}}\hat{a}_{\mathrm{s},\mathrm{in}}(t), \nonumber
\end{eqnarray}
%
where $\gamma_{j}$ are the rates of exchange of excitations of the modes with
their corresponding environments, and the input operators have zero
mean, $\langle\hat{a}_{j,\mathrm{in}}(t)\rangle=\langle\hat{p}_\mathrm{m,in}(t)\rangle=0$), and non-zero two-time correlators%
\begin{equation}
\langle\hat{a}_{j,\mathrm{in}}(t)\hat{a}_{j,\mathrm{in}}^{\dag}(t^{\prime
})\rangle=\langle\hat{p}_{\mathrm{m,in}}(t)\hat{p}_{\mathrm{m,in}}(t^{\prime
})\rangle=\delta(t-t^{\prime}), \label{InputCorr}%
\end{equation}
and play the role of the environmental quantum fluctuations driving the
system. In this equations we have assumed to be working in the
high-temperature limit where the number of phonons at thermal equilibrium can
be approximated by $\bar{n}_{\mathrm{th}}\approx k_{B}T/\hbar\Omega
_{\mathrm{m}}\gg1$.

Before studying the equations, it is convenient to make a variable
change that will allow us to see how many free parameters they really have. To
this aim, we define the following normalized parameters%
\begin{eqnarray}
g_{\mathrm{DC}}&=&\frac{\chi}{\sqrt{\gamma_{\mathrm{p}}\gamma_{\mathrm{s}}}},\hspace{3mm}\sigma=\frac{\chi\mathcal{E}_{\mathrm{L}}}{\gamma_{\mathrm{p}%
}\gamma_{\mathrm{s}}},\hspace{3mm}\kappa=\frac{\gamma_{\mathrm{p}}}%
{\gamma_{\mathrm{s}}},
\\
\delta_{j}&=&\frac{\Delta_{j}}{\gamma_{j}}\hspace{3mm}\gamma=\frac{\gamma_{\mathrm{m}}}{\gamma_{\mathrm{s}}},\hspace{3mm}\Omega=\frac{\Omega_{\mathrm{m}}}{\gamma_{\mathrm{s}}},\hspace{3mm}g=\frac{g_{\mathrm{s}}/\gamma_{\mathrm{s}}}{g_{\mathrm{DC}}\sqrt{\Omega}}, \nonumber
\end{eqnarray}
time $\tau=\gamma_{\mathrm{s}}t$, system operators
\begin{eqnarray}\label{NormVars}
\hat{b}_{\mathrm{s}}=g_{\mathrm{DC}}\hat{a}_{\mathrm{s}},\hspace{3mm}\hat
{b}_{\mathrm{p}}=\sqrt{\kappa}g_\mathrm{DC}\hat{a}_{\mathrm{p}},
\\
\hat{p}=\frac{g_{\mathrm{DC}}}{\sqrt{\Omega}}\hat{p}_{\mathrm{m}},\hspace{3mm}\hat
{x}=g_{\mathrm{DC}}\sqrt{\Omega}\hat{x}_{\mathrm{m}},\nonumber
\end{eqnarray}
and input operators
\begin{eqnarray}
\hat{b}_{j,\mathrm{in}}(\tau)  =\frac{1}{\sqrt{\gamma_{\mathrm{s}}}}\hat
{a}_{j,\mathrm{in}}(\tau/\gamma_{\mathrm{s}}),
\\
\hat{p}_{\mathrm{in}}(\tau)  =\frac{1}{\sqrt{\gamma_{\mathrm{s}}}}\hat
{p}_{\mathrm{m,in}}(\tau/\gamma_{\mathrm{s}}), \nonumber
\end{eqnarray}
which satisfy the same correlators as before, see Eq. (\ref{InputCorr}), but now
with respect to the dimensionless time $\tau$. With these changes, the quantum
Langevin equations are transformed into
\begin{eqnarray}\label{LangNormEqs}
\frac{d\hat{x}}{d\tau}  &=& \Omega^{2}\hat{p},
\\
\frac{d\hat{p}}{d\tau}  &=& -\gamma\hat{p}-\hat{x}+2g\hat{b}_{\mathrm{s}%
}^{\dagger}\hat{b}_{\mathrm{s}}+\sqrt{\frac{4\gamma\bar{n}_{\mathrm{th}}%
}{\Omega}}g_{\mathrm{DC}}\hat{p}_{\mathrm{in}}(\tau),\nonumber
\\
\frac{1}{\kappa}\frac{d\hat{b}_{\mathrm{p}}}{d\tau}  &=& \sigma-(1-\mathrm{i}%
\delta_{\mathrm{p}})\hat{b}_{\mathrm{p}}-\frac{1}{2}\hat{b}_{\mathrm{s}}%
^{2}+\sqrt{2}g_{\mathrm{DC}}\hat{b}_{\mathrm{p},\mathrm{in}}(t),\nonumber
\\
\frac{d\hat{b}_{\mathrm{s}}}{d\tau}  &=& -(1-\mathrm{i}\delta_{\mathrm{s}%
}-\mathrm{i}g\hat{x})\hat{b}_{\mathrm{s}}+\hat{b}_{\mathrm{p}}\hat
{b}_{\mathrm{s}}^{\dag}+\sqrt{2}g_{\mathrm{DC}}\hat{b}_{\mathrm{s}%
,\mathrm{in}}(t).\nonumber
\end{eqnarray}
It is interesting to note that $g$ basically provides the ratio between the
single-photon optomechanical and down-conversion couplings, and hence,
assuming $\sqrt{\Omega}$ to be of order 1, they inform us about which of the two
nonlinear processes dominates.

The quantum langevin equations are nonlinear operator equations, and hence it
is a formidable task to obtain results directly from them without further
approximations. In the following, we analyze two relevant limits of these
equations: the classical limit and the limit of `small' quantum fluctuations.

\section{Classical analysis}

The classical limit of the model is obtained by assuming that all the modes of
the system are in a coherent state. Defining the corresponding amplitudes
$\beta_{j}=\langle\hat{b}_{j}\rangle$, $x=\langle\hat{x}\rangle$, and
$p=\langle\hat{p}\rangle$, and taking the expectation value of the quantum
Langevin equations (\ref{LangNormEqs}), such approximation leads to the
classical equations
\begin{eqnarray}\label{NormEqs}
\dot{x} &=& \Omega^{2}p,
\\
\dot{p} &=& -\gamma p-x+2g\left\vert \beta_{\mathrm{s}}\right\vert
^{2},\nonumber
\\
\kappa^{-1}\dot{\beta}_{\mathrm{p}}  &=& \sigma-(1-\mathrm{i}\delta
_{\mathrm{p}})\beta_{\mathrm{p}}-\beta_{\mathrm{s}}^{2}/2,\nonumber
\\
\dot{\beta}_{\mathrm{s}}  &=& -(1-\mathrm{i}\delta_{\mathrm{s}}-\mathrm{i}%
gx)\beta_{\mathrm{s}}+\beta_{\mathrm{p}}\beta_{\mathrm{s}}^{\ast}.\nonumber
\end{eqnarray}
As non-trivial nonlinear equations, it is not possible to find their
time-dependent analytical solutions other than numerically. However, working
with a dissipative system, we are mainly interested in its behavior for long
times (asymptotic limit), and there is a lot that we can say about this
without really solving the full nonlinear equations; in particular, we follow
closely the procedure already applied to detuned DOPOs
\cite{Lugiato88,Pettiaux89,Fabre90}.

\begin{figure*}[t]
\includegraphics[width=\textwidth]{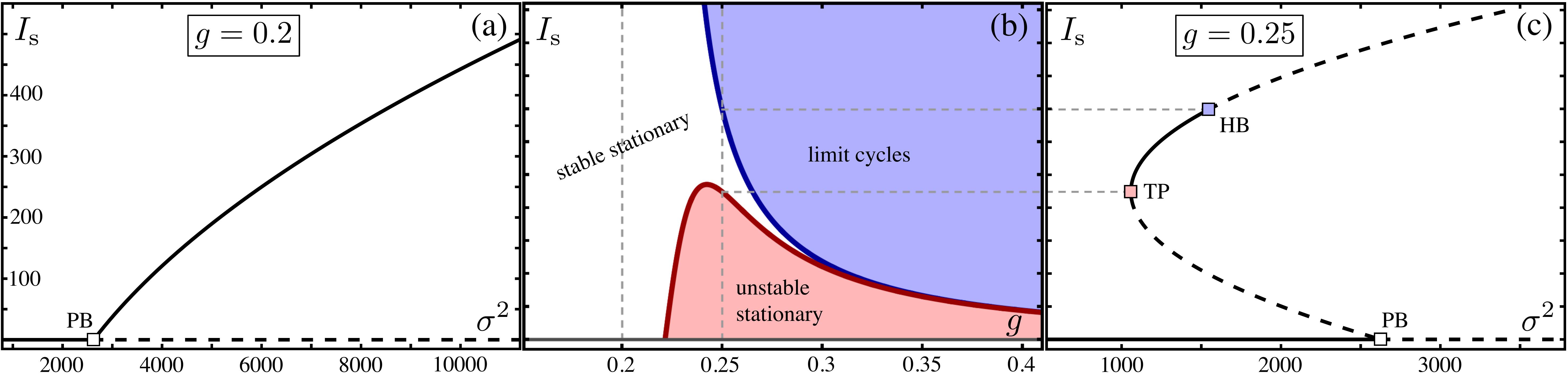}\caption{We show the
bifurcations and corresponding stable and unstable regions for one particular
example for which we have chosen $\gamma=0.005$, $\Omega=-\delta_{\mathrm{s}%
}=10$, $\delta_{\mathrm{p}}=5$, and $\kappa=100$, the first three being
typical parameters when aiming for sideband cooling in optomechanical systems.
Note that this fixes all the parameters but $g$ and
$I_{\mathrm{s}}$ (we are using the steady-state intensity $I_{\mathrm{s}}$ as
a parameter instead of $\sigma$, because the latter can be uniquely determined
from the former, but in general not the other way around). In (b) we show in
the space of these parameters the turning point (red thick line) and the
single Hopf instability (blue thick line) found in this example, coloring the
regions where they make the stationary solution unstable. Note that for this
choice of parameters, in the absence of optomechanical coupling there are no
instabilities apart from the trivial pitchfork bifurcation, as can be checked
from conditions (\ref{TPcond}) and (\ref{HBcond}). Hence, we see that the
effect of the optomechanical coupling in this case consists in introducing new
instabilities which greatly reduce the domain of stabity of the nontrivial
stationary state. In (a) and (c) we show how the steady-state intensity
$I_{\mathrm{s}}$ depends on the injection $\sigma^{2}$ for two specific values
of $g$, corresponding to the vertical grey dashed lines in the
parameter space (b), denoting its unstable and stable regions by dashed and
solid lines, respectively. In (a) we have chosen $g=0.2$, for
which no instabilities are present in the nontrivial solution as can be
appreciated in (b), and hence only the pitchfork instability connecting the
trivial and nontrivial solutions is present. In (c), on the other hand, we
have chosen $g=0.25$, for which we find both a turning point (and
hence a domain of bistability between the trivial and nontrivial solutions)
and a Hopf bifurcation leading to time-dependent asymptotic solutions.}%
\label{Fig2}%
\end{figure*}

\subsection{Stationary solutions}

The simplest asymptotic behavior that one can expect is that the system
reaches some steady state. Hence, it is always convenient to start by finding
the time-independent solutions to the nonlinear equations, which we denote by
a bar, e.g., $\bar{x}$; when needed, we will write the complex field
amplitudes as $\bar{\beta}_{j}=\sqrt{I_{j}}\exp(\mathrm{i}\varphi_{j})$, with
real variables $\varphi_{j}\in]-\pi,\pi]$ and $I_{j}\geq0$.

All the stationary solutions of (\ref{NormEqs}) have $\bar{p}=0$ and $\bar
{x}=2gI_{\mathrm{s}}$, leaving us with
\begin{subequations}
\label{SSfields}%
\begin{align}
\sigma &  =(1-\mathrm{i}\delta_{\mathrm{p}})\bar{\beta}_{\mathrm{p}}%
+\bar{\beta}_{\mathrm{s}}^{2}/2,\label{SSp}\\
\bar{\beta}_{\mathrm{p}}\bar{\beta}_{\mathrm{s}}^{\ast}  &  =(1-\mathrm{i}%
\delta_{\mathrm{s}}-2\mathrm{i}g^{2}I_{\mathrm{s}})\bar{\beta}_{\mathrm{s}}.
\label{SSs}%
\end{align}
We distinguish then two types of stationary solutions: \textit{trivial} or
\textit{below-threshold} solutions, which have $I_{\mathrm{s}}=0$, and
\textit{nontrivial} or \textit{above-threshold} solutions with $I_{\mathrm{s}%
}\neq0$.

In the trivial case, the solution is simply
\end{subequations}
\begin{equation}
\bar{\beta}_{\mathrm{s}}=0\text{ \ \ and \ \ }\bar{\beta}_{\mathrm{p}}%
=\sigma/(1-\mathrm{i}\delta_{\mathrm{p}})\text{.} \label{TrivialSol}%
\end{equation}
As for the nontrivial solutions, we find their analytic expression as follows.
First, note that (\ref{SSs}) implies $\bar{\beta}_{\mathrm{p}}=(1-\mathrm{i}\delta_{\mathrm{s}}-2\mathrm{i}%
g^{2}I_{\mathrm{s}})e^{2\mathrm{i}\varphi_{\mathrm{s}}}$, which plugged into (\ref{SSp}) leads to $e^{-2\mathrm{i}\varphi_{\mathrm{s}}}\sigma=(1-\mathrm{i}\delta_{\mathrm{p}})(1-\mathrm{i}\delta_{\mathrm{s}}-2\mathrm{i}g^{2}I_{\mathrm{s}})+I_{\mathrm{s}}/2$, whose absolute value squared gives us a second order polynomial for the signal
intensity%
\begin{eqnarray}
\sigma^{2}&=&\left[  1+\frac{I_{\mathrm{s}}}{2}-\delta_{\mathrm{p}}%
(\delta_{\mathrm{s}}+2g^{2}I_{\mathrm{s}})\right]^{2}+(\delta_{\mathrm{p}%
}+\delta_{\mathrm{s}}+2g^{2}I_{\mathrm{s}})^{2}\nonumber
\\
&\equiv& q_{0}+q_{1}%
I_{\mathrm{s}}+q_{2}I_{\mathrm{s}}^{2}, \label{SteadyStateEquationIs}
\end{eqnarray}
with
\begin{subequations}\label{qcoefs}
\begin{eqnarray}
q_{0}&=&(1+\delta_{\mathrm{p}}^{2})(1+\delta_{\mathrm{s}}^{2}),
\\
q_{1}&=&(1-\delta_{\mathrm{p}}\delta_{\mathrm{s}})+4\delta_{\mathrm{s}}(1+\delta_{\mathrm{p}}^{2})g^{2},
\\
q_{2}&=&4g^{4}+\left(\frac{1}{2}-2\delta_{\mathrm{p}}g^{2}\right)  ^{2}.
\end{eqnarray}
\end{subequations}
Depending on the value of the parameters, this equation can have a single real
positive solution or two, as shown in Fig. \ref{Fig2}. In order to find for
which values of the parameters (in particular of the injection $\sigma$ and
the detunings $\delta_{j}$, experimentally tunable) this happens, we just need to obtain the
expression for the turning point, marked as TP in Fig. \ref{Fig2}, which is
nothing but the extremum of $\sigma^{2}(I_{\mathrm{s}})$, that is,
\begin{equation}
\left.  \frac{\partial\sigma^{2}}{\partial I_{\mathrm{s}}}\right\vert
_{I_{\mathrm{s}}=I_{\mathrm{s}}^{\mathrm{TP}}}=0\hspace{3mm}\Longrightarrow
\hspace{3mm}I_{\mathrm{s}}^{\mathrm{TP}}=-q_{1}/2q_{2}.
\end{equation}
Taking into account that $q_{2}>0$, the turning point will exist only if
$q_{1}<0$, which gives us a condition on the detunings for a given
optomechanical coupling:%
\begin{equation}
\delta_{\mathrm{p}}\delta_{\mathrm{s}}>1+4\delta_{\mathrm{s}}(1+\delta
_{\mathrm{p}}^{2})g^{2}. \label{TPcond}%
\end{equation}
Hence, when this condition is satisfied, we will have two possible
steady-state signal intensities (three counting the trivial one) for
injections $\sigma^{2}\in]q_{0}-q_{1}^{2}/4q_{2},q_{0}]$, see Fig. \ref{Fig2}c. Let us anticipate,
however, that the branch connecting the trivial solution with the upper branch
of the nontrivial one is unstable (see Fig. \ref{Fig2}), so only two out of
the three possible solutions can be observed in real experiments, leading to an intensity
bistability common in nonlinear optical systems. Finally, notice that
condition (\ref{TPcond}) with $g=0$ is in agreement with that found for
detuned DOPOs \cite{Lugiato88,Pettiaux89,Fabre90}.

\subsection{Linear stability analysis}\label{LinStabAn}

The existence of a mathematical solution of the nonlinear equations is not
enough to ensure its physical reality: it also needs to be stable against
perturbations, since in the real world these are unavoidable, and therefore we
would never be able to observe the system in the corresponding solution
otherwise. Hence, in the following we proceed to study the stability of the
stationary solutions found above.

Let us collect the variables of the system in a vector $\mathbf{b}%
=\operatorname{col}(x,p,\beta_{\mathrm{p}},\beta_{\mathrm{p}}^{\ast}%
,\beta_{\mathrm{s}},\beta_{\mathrm{s}}^{\ast})$. The stability of a given
stationary solution $\mathbf{\bar{b}}$\ can be analyzed as follows
\cite{Narducci88}. We consider small fluctuations around it by writing
$\mathbf{b}(t)=\mathbf{\bar{b}}+\delta\mathbf{b}(t)$, introduce this ansatz
into the nonlinear system (\ref{NormEqs}), and keep only terms which are
linear in the fluctuations, obtaining a linear system $\delta\mathbf{\dot{b}%
}=\mathcal{L}\delta\mathbf{b}$, where $\mathcal{L}$ is the so-called
\textit{linear stability matrix}. This matrix depends on the system parameters
and the particular stationary solution whose stability we are considering, and
in our case is given by
\begin{widetext}
\begin{equation}
\mathcal{L}=\left(
\begin{array}
[c]{cccccc}%
0 & \Omega^{2} & 0 & 0 & 0 & 0\\
-1 & -\gamma & 0 & 0 & 2g\bar{\beta}_{\mathrm{s}}^{\ast} & 2g\bar{\beta
}_{\mathrm{s}}\\
0 & 0 & -\kappa(1-\mathrm{i}\delta_{\mathrm{p}}) & 0 & -\kappa\bar{\beta
}_{\mathrm{s}} & 0\\
0 & 0 & 0 & -\kappa(1+\mathrm{i}\delta_{\mathrm{p}}) & 0 & -\kappa\bar{\beta
}_{\mathrm{s}}^{\ast}\\
\mathrm{i}g\bar{\beta}_{\mathrm{s}} & 0 & \bar{\beta}_{\mathrm{s}}^{\ast} &
0 & -(1-\mathrm{i}\delta_{\mathrm{s}}-\mathrm{i}g\bar{x}) & \bar{\beta
}_{\mathrm{p}}\\
-\mathrm{i}g\bar{\beta}_{\mathrm{s}}^{\ast} & 0 & 0 & \bar{\beta}_{\mathrm{s}}
& \bar{\beta}_{\mathrm{p}}^{\ast} & -(1+\mathrm{i}\delta_{\mathrm{s}%
}+\mathrm{i}g\bar{x})
\end{array}
\right)  . \label{Lmatrix}
\end{equation}
\end{widetext}
Since the equation for the fluctuations is linear, it is then clear that their
dynamical behaviour is controlled by the eigenvalues of this matrix. In particular, the
fluctuations will be damped and disappear in the asymptotic limit only if the
real part of all the eigenvalues is negative. Hence, we say that a stationary
solution $\mathbf{\bar{b}}$ is stable (and therefore physical) when all the
eigenvalues of its corresponding linear stability matrix $\mathcal{L}%
(\mathbf{\bar{b}})$ have negative real part.

The points in the parameter space where at least one of the eigenvalues has
zero real part are known as \textit{critical points}, \textit{instabilities},
or \textit{bifurcations}, and they separate the regions in which the
stationary solution changes from stable to unstable. We can distinguish two
types of instabilities: \textit{pitchfork} or \textit{static} bifurcations,
where the imaginary part of the relevant eigenvalue is also zero, which
connect the stationary solution with another stationary solution; and
\textit{Hopf} or \textit{dynamic} bifurcations, where the imaginary part of
the relevant eigenvalue is non-zero, which connect the stationary solution
with a time-dependent solution (usually some periodic solution, known in this
context as a \textit{periodic orbit }or\textit{ limit cycle}).

Before proceeding, let us comment on one subtle point concerning the system
parameters. The linear stability matrix (\ref{Lmatrix}) does not depend
explicitly on the injection $\sigma$, it does only implicitly through the
intracavity stationary amplitudes $\bar{\beta}_{\mathrm{p}}$ and $\bar{\beta
}_{\mathrm{s}}$. It is then convenient to use either $I_{\mathrm{p}}$ or
$I_{\mathrm{s}}$ as a parameter instead of $\sigma$ when dealing with the
trivial or nontrivial solutions, respectively, knowing that $\sigma$ can
always be uniquely determined from them by using (\ref{TrivialSol}) or
(\ref{SteadyStateEquationIs}).

\btxtb{Let us now proceed to discuss the instabilities that can be found in the DOMPO. We provide here a summary of the main results, and leave the detailed derivations for Appendix \ref{StabApp}. Concerning the trivial solution, it possess only one bifurcation appearing when $I_{\mathrm{p}}=1+\delta_{\mathrm{s}}^{2}$. The trivial solution becomes unstable for $I_{\mathrm{p}}>1+\delta_{\mathrm{s}}^{2}$, or in terms of the injection, when $\sigma^{2}>(1+\delta_{\mathrm{s}}^{2})(1+\delta_{\mathrm{p}}^{2})$. Note that this is precisely the point at which the trivial and nontrivial solutions coalesce, see the points marked as PB in Fig. \ref{Fig2}, and hence this pitchfork bifurcation simply connects these two stationary solutions. As for the nontrivial solution, it provides one more static instability at $I_{\mathrm{s}}=I_{\mathrm{s}}^{\mathrm{TP}}$. Hence, we see that the turning point of the nontrivial solution is an instability, and it is simple
to check that the lower branch of the nontrivial solution connecting the upper branch with the trivial solution is unstable (for example by evaluating the
eigenvalues numerically for one set of parameters), as shown in Fig. \ref{Fig2}. In other words, the turning point is a pitchfork bifurcation connecting the unstable lower branch with the upper branch, which is stable in all its domain of existence, except for possible Hopf bifurcations which we discuss next.

The behaviour of the dynamical instabilities of the nontrivial solution is very rich in the DOMPO. Let us first note that in the absence of optomechanical coupling ($g=0$) there is a single Hopf bifurcation located at (see Appendix \ref{StabApp} and \cite{Lugiato88,Pettiaux89,Fabre90})
\begin{equation}\label{IpHB}
I_{\mathrm{s}}^{\mathrm{HB}}=-\frac{(1+\delta_{\mathrm{p}}^{2})[(2+\kappa)^{2}+\kappa^{2}\delta_{\mathrm{p}}^{2}]}{(1+\kappa)^{2}(2+\kappa+\kappa
\delta_{\mathrm{p}}^{2}+2\delta_{\mathrm{p}}\delta_{\mathrm{s}})},
\end{equation}
which requires
\begin{equation}
\delta_{\mathrm{p}}\delta_{\mathrm{s}}<-1-\kappa(1+\delta_{\mathrm{p}}^{2})/2, \label{HBcond}
\end{equation}
to exist (otherwise $I_{\mathrm{s}}^{\mathrm{HB}}<0$), which incidentally means that it does not exist when there is bistability in the system (what
requires $\delta_{\mathrm{p}}\delta_{\mathrm{s}}>1$). It is possible to show that the portion of the nontrivial solution with $I_{\mathrm{s}}>I_{\mathrm{s}}^{\mathrm{HB}}$ becomes unstable, and the limit cycles become chaotic for large enough injections \cite{Lugiato88,Pettiaux89,Fabre90}. The main effect of optomechanics, that is, of increasing $g$, is both changing the location of this dynamical instability already present for $g=0$, as well as creating new ones that cannot be understood as a deformation of the latter. This is what we show in Fig. \ref{Fig2}b for one example, where we plot the signal intensity of the Hopf instability that we have found as a function of $g$. Let us remark that in the $g\neq0$ case the complicated form of the linear stability matrix has prevented us from finding simple analytic expressions for such instabilities, but we have been able to find a simple way to compute them efficiently with the help of symbolic programs that has allowed us to perform an exhaustive analysis, see Appendix \ref{StabApp}.}

\section{Quantum properties within the linearized description}

In order to characterize the quantum properties of the DOMPO, we now apply the
widely-used method of standard linearization
\cite{Drummond80,Lugiato81,Collett84,Navarrete14}. In this approach, one
assumes that the asymptotic classical solution of the system is a strong
attractor, and hence quantum mechanics acts just as strongly damped
fluctuations or noise driving continuously the system and trying to bring it
out of equilibrium. As we did in the classical analysis, let us collect the
fundamental operators of the system into a vector $\mathbf{\hat{b}%
}=\operatorname{col}(\hat{x},\hat{p},\hat{b}_{\mathrm{p}},\hat{b}_{\mathrm{p}%
}^{\dagger},\hat{b}_{\mathrm{s}},\hat{b}_{\mathrm{s}}^{\dagger})$. In this
scenario, it is then convenient to write the operators as $\mathbf{\hat{b}%
}(\tau)=\mathbf{\bar{b}}+\delta\mathbf{\hat{b}}(\tau)$, where we remind that
$\mathbf{\bar{b}}$ is the stationary solution found for the operators within
the classical description of the system (we will not consider periodic orbits
in this work, although they can also be studied with a similar approach).
Assuming that $\mathbf{\bar{b}}$ is a strong classical attractor, that is,
that the eigenvalues of its associated linear stability matrix $\mathcal{L}$
are large enough for quantum fluctuations to be strongly damped, one can
neglect terms of the quantum Langevin equations (\ref{LangNormEqs}) beyond linear order in the fluctuations $\delta\mathbf{\hat{b}}$, turning them into the
linear system
\begin{equation}
\frac{d}{d\tau}\delta\mathbf{\hat{b}}=\mathcal{L}\delta\mathbf{\hat{b}}%
+\sqrt{2}g_{\mathrm{DC}}\widehat{\mathbf{f}}(\tau), \label{LinLan}%
\end{equation}
where we have defined the input-vector operator
\begin{equation}
\widehat{\mathbf{f}}(\tau)=\operatorname{col}\left(  0,\sqrt{\frac{2\gamma\bar
{n}_{\mathrm{th}}}{\Omega}}\hat{p}_{\mathrm{in}},\hat{b}_{\mathrm{p,in}}%
,\hat{b}_{\mathrm{p,in}}^{\dagger},\hat{b}_{\mathrm{s,in}},\hat{b}%
_{\mathrm{s,in}}^{\dagger}\right)  , \label{f}%
\end{equation}
which acts precisely as a quantum force continuously driving the system out of equilibrium. Note that the two-time input correlators can be written in the compact form $\langle\hat{f}_{j}(\tau)\hat{f}_{l}(\tau^{\prime})\rangle=M_{jl}\delta(\tau-\tau^{\prime})$, where $M_{jl}$ are the elements of the matrix $\mathcal{M}=\mathcal{M}_{\mathrm{m}}\oplus\mathcal{M}_{\mathrm{p}}\oplus\mathcal{M}_{\mathrm{s}}$
with
\begin{eqnarray}\label{Mmatrix}
\mathcal{M}_{\mathrm{m}}&=&\left(
\begin{array}
[c]{cc}%
0 & 0\\
0 & 2\gamma\bar{n}_{\mathrm{th}}/\Omega
\end{array}
\right),
\\
\mathcal{M}_{\mathrm{p}}&=&\left(
\begin{array}
[c]{cc}%
0 & \kappa\\
0 & 0
\end{array}
\right),\nonumber
\\
\mathcal{M}_{\mathrm{s}}&=&\left(
\begin{array}
[c]{cc}%
0 & 1\\
0 & 0
\end{array}
\right). \nonumber
\end{eqnarray}

In the following, we will particularize these linearized quantum Langevin
equations to the two types of classical stable stationary solutions that we have
found (trivial and nontrivial), analyzing the behavior that they predict for
the mechanical mode. From the previous discussion, it is clear that such a
linearized description will break down close to the critical points of the
classical theory, but as proven again and again in many nonlinear optical
systems, its predictions usually provide the correct tendency of observables
as one approaches the critical points, at the very least qualitatively. There
are however some exceptions, corresponding to cases in which the quantum
mechanical effects are purely nonlinear, so that linearization completely
misses them. Indeed, we shall see one of such examples now.

\subsection{Failure of the method below threshold}

The first interesting thing to note about the linearization approach is how
it completely fails to capture any optomechanical phenomena that might be
occurring below threshold, that is, when the signal field is switched off
classically, so that there is no coherent background in the mechanical and
signal modes. This is clearly seen from the fact that, as we show in Sec.
\ref{BTstab} of Appendix \ref{StabApp}, the linear stability matrix is written as a direct sum of
matrices acting on each mode, so that the equations for the mechanical
fluctuations $\delta\hat{x}=\hat{x}$ and $\delta\hat{p}=\hat{p}$ simply read
\begin{subequations}
\label{BTmechEqs}%
\begin{align}
\frac{d\hat{x}}{d\tau}  &  =\Omega^{2}\hat{p},\\
\frac{d\hat{p}}{d\tau}  &  =-\gamma\hat{p}-\hat{x}+\sqrt{\frac{4\gamma\bar
{n}_{\mathrm{th}}}{\Omega}}g_{\mathrm{DC}}\hat{p}_{\mathrm{in}}(\tau).
\end{align}
These equations receive absolutely no information from the optical modes, in
particular corresponding to a harmonic oscillator in thermal equilibrium with
its environment. In the same way, the fluctuations of the optical mode
$\delta\hat{b}_{\mathrm{s}}=\hat{b}_{\mathrm{s}}$ are completely unaffected by
the mechanics, since they obey the usual below-threshold DOPO dynamics,%
\end{subequations}
\begin{equation}
\frac{d\hat{b}_{\mathrm{s}}}{d\tau}=(-1+\mathrm{i}\delta_{\mathrm{s}})\hat
{b}_{\mathrm{s}}+\bar{\beta}_{\mathrm{p}}\hat{b}_{\mathrm{s}}^{\dagger}%
+\sqrt{2}g_{\mathrm{DC}}\hat{b}_{\mathrm{s,in}}(\tau). \label{BTsignalEqs}%
\end{equation}

As is well known \cite{BlueBook,WMbook,NavarretePhDthesis}, the latter
equations predict that the signal mode gets more and more squeezed as the
critical point of the below threshold solution $I_{\mathrm{p}}=1+\delta
_{\mathrm{s}}^{2}$ is approached, \cnb{denoted by} PB in Figs. \ref{Fig2}a and \ref{Fig2}c. In
fact, the linearized description predicts an infinite photon number exactly at
threshold \cite{Drummond80,Lugiato81,Collett84}, which of course gets
regularized once more accurate approaches are used
\cite{Navarrete14,Kinsler93,Kinsler95,Drummond02,Chaturvedi99,Pope00,MertensPRL93,MertensPRA93,Veits97,Degenfeld15}. In the light of this insight, it is hard to believe that despite the large number of
photons \cnb{present} in the signal mode, optomechanical scattering will have no
effect whatsoever on the fluctuations of the modes, as\ Eqs. (\ref{BTmechEqs})
and (\ref{BTsignalEqs}) predict, and one has to conclude that the
linearization simply fails to capture whatever phenomena occurs below
threshold. In fact, in this regime the signal photons scattered by the
mechanical mode are purely quantum mechanical, with no coherent or classical
background, and this is precisely what makes the optomechanical interaction
$\hat{a}_{\mathrm{s}}^{\dagger}\hat{a}_{\mathrm{s}}\hat{x}_{\mathrm{m}}$
purely nonlinear or \textit{nongaussian}, which is ultimately the reason why
any effect related to it is completely lost upon linearization.

More sophisticated linearized descriptions such as the self-consistent
linearization introduced in \cite{Navarrete14} cannot work either, since it is
equivalent to making a Gaussian ansatz for the full state of the system, while
the optomechanical interaction becomes purely nongaussian below threshold as
we have argued above. Hence, more elaborated techniques are required below
threshold, such as a numerical simulation based on \rtxt{the positive P
representation \cite{PositiveP,Kinsler93,Kinsler95,Drummond02,CarmichaelBook2} or the self-consistent Mori Projector operator (c-MoP) theory
\cite{Degenfeld14,Degenfeld15}}, which are beyond the scope of this work, and we indeed explore in other works \cite{Degenfeld15b}.
Nevertheless, as we are about to see, standard linearization can still be a
useful tool allowing us to analyze the system above threshold, and find
indications of interesting phenomena.

\subsection{Above-threshold predictions}

The situation is rather different above threshold. In this case all the modes
have a classical background, and hence it is possible to linearize the
optomechanical interaction without loosing it completely. Consequently, it is
to be expected that, even though the method will fail at the critical points
of the classical theory, it will provide us with a good qualitative picture of
the trend that the quantum properties of the system follow.

\begin{figure*}[t]
\includegraphics[width=0.7\textwidth]{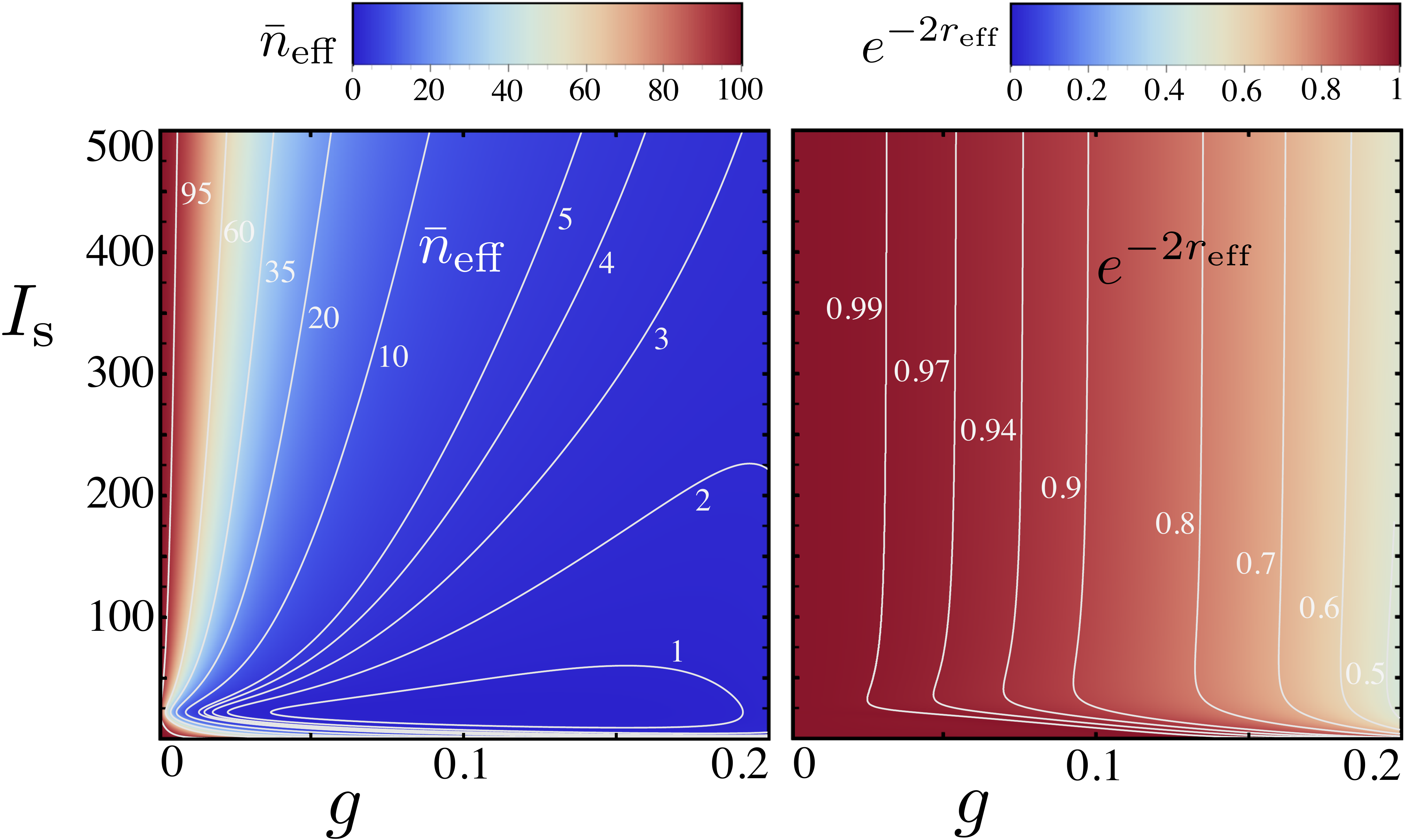}\caption{Density plots
of the effective phonon number $\bar{n}_{\mathrm{eff}}$ and mechanical
squeezing $e^{-2r_{\mathrm{eff}}}$ as a function of the optomechanical
coupling $g$ and the intracavity signal intensity $I_{\mathrm{s}}$. The rest
of parameters have the same values as in Fig. \ref{Fig2}, and we assume a
phonon number $\bar{n}_{\mathrm{th}}=100$ at thermal equilibrium. Note how the
mechanical state is strongly affected by the optomechanical interaction, which
allows both for cooling and squeezing the mechanical motion.}%
\label{FigResultsGs}%
\end{figure*}

As explained above, the linearization method is equivalent to making a
Gaussian ansatz \cite{Braunstein05, Weedbrook12,NavarreteQI} for the \gtxt{state of
the quantum fluctuations around the classical solution
\cite{Navarrete14,Degenfeld15}}. Hence, within this approach the quantum
properties of the system are completely characterized by the covariance matrix
of all the modes \cite{Braunstein05, Weedbrook12,NavarreteQI}. Given the position $\hat{x}_{j}=\hat{a}_{j}^{\dagger}+\hat{a}_{j}$ and
momentum $\hat{p}_{j}=\mathrm{i}(\hat{a}_{j}^{\dagger}-\hat{a}_{j})$
quadratures of the optical modes ($j=\mathrm{p},\mathrm{s}$), and defining the
quadrature vector operator $\mathbf{\hat{r}}=\operatorname{col}(\hat
{x}_{\mathrm{m}},\hat{p}_{\mathrm{m}},\hat{x}_{\mathrm{p}},\hat{p}%
_{\mathrm{p}},\hat{x}_{\mathrm{s}},\hat{p}_{\mathrm{s}})$, the covariance
matrix $V$ is defined as the symmetric matrix with elements $V_{jl}%
=\langle\delta\hat{r}_{j}\delta\hat{r}_{l}+\delta\hat{r}_{l}\delta\hat{r}%
_{j}\rangle/2$ \cite{Braunstein05, Weedbrook12,NavarreteQI}. \btxtb{In Appendix \ref{CovApp} we explain how this object can be efficiently evaluated numerically for any value of the parameters directly from the linearized Langevin equations (\ref{LinLan}), specifically from the eigensystem of the linear stability matrix (\ref{Lmatrix}).}

On the other hand, as explained in the introduction, the main question that we would like to
explore is the effect that the optomechanical interaction has on the
mechanical state. In the following we show through a set of examples how the squeezed down-converted field is able to cool down the
mechanical motion. \cnb{Moreover,} it does so in a way that can be much less sensitive to
parameters than standard sideband cooling. \cnb{Furthermore}, apart from cooling it,
the example will show a trend of the optomechanical interaction to squeeze the
thermal mechanical motion.

In order to show this, we first find the reduced state of the mechanical mode.
Since within the linearized picture the state of the whole system is Gaussian,
the reduced mechanical state is Gaussian as well, with a covariance matrix
given by the corresponding submatrix of the full covariance matrix
(\ref{CovMat}) \cite{NavarreteQI}:%
\begin{equation}
V_{\mathrm{m}}=\left(
\begin{array}
[c]{cc}%
V_{11} & V_{12}\\
V_{21} & V_{22}%
\end{array}
\right)  .
\end{equation}
\cnb{In order to get a better physical picture of the mechanical state}, we further exploit the fact that any single-mode Gaussian state can be written as a
squeezed thermal state up to a rotation in phase space
\cite{Weedbrook12,NavarreteQI} (which simply provides the direction of phase space along which squeezing occurs). This means that the mechanical state
$\hat{\rho}_{\mathrm{m}}$ can be written in the form%
\begin{equation}
\hat{R}(\theta)\hat{\rho}_{\mathrm{m}}\hat{R}^{\dagger}(\theta)=\hat
{S}(r_{\mathrm{eff}})\hat{\rho}_{\mathrm{th}}(\bar{n}_{\mathrm{eff}})\hat
{S}^{\dagger}(r_{\mathrm{eff}})\equiv\tilde{\rho}_{\mathrm{m}}, \label{RhoEff}%
\end{equation}
for some phase-shift operator $\hat{R}(\theta)=\exp[\mathrm{i}\theta(\hat
{x}_{\cnb{\mathrm{m}}}^{2}+\hat{p}_{\cnb{\mathrm{m}}}^{2})/4]$, where $\hat{\rho}_{\mathrm{th}}(\bar{n}%
_{\mathrm{eff}})$ is a thermal state with effective phonon number $\bar
{n}_{\mathrm{eff}}$, and $\hat{S}(r_{\mathrm{eff}})=\exp[\mathrm{i}r_{\mathrm{eff}%
}\hat{x}_\cnb{\mathrm{m}}\hat{p}_\cnb{\mathrm{m}}/2]$ is a squeezing operator with effective squeezing parameter $r_\mathrm{eff}$. We are in particular interested
in the effective phonon number $\bar{n}_{\mathrm{eff}}$ and the effective
mechanical squeezing $e^{-2r_{\mathrm{eff}}}$, \btxtb{which can be written in terms of the elements of the covariance matrix as (see Appendix \ref{CovApp})
\begin{equation}\label{EffPar}
\bar{n}_{\mathrm{eff}}=\left(  \sqrt{V_{+}V_{-}}-1\right)/2 \hspace{2mm}\text{and}\hspace{2mm} e^{-2r_{\mathrm{eff}}}=\sqrt{V_{-}/V_{+}},%
\end{equation}
where $V_{\mp}=\mathrm{tr}\{V_{\mathrm{m}}\}/2\mp\sqrt{\mathrm{tr}\{V_{\mathrm{m}}\}^{2}/4-\det\{V_{\mathrm{m}}\}}$ are the eigenvalues of the mechanical covariance matrix. Let us now pass to analyze these parameters for some specific situation.}
\begin{figure*}[t]
\includegraphics[width=0.7\textwidth]{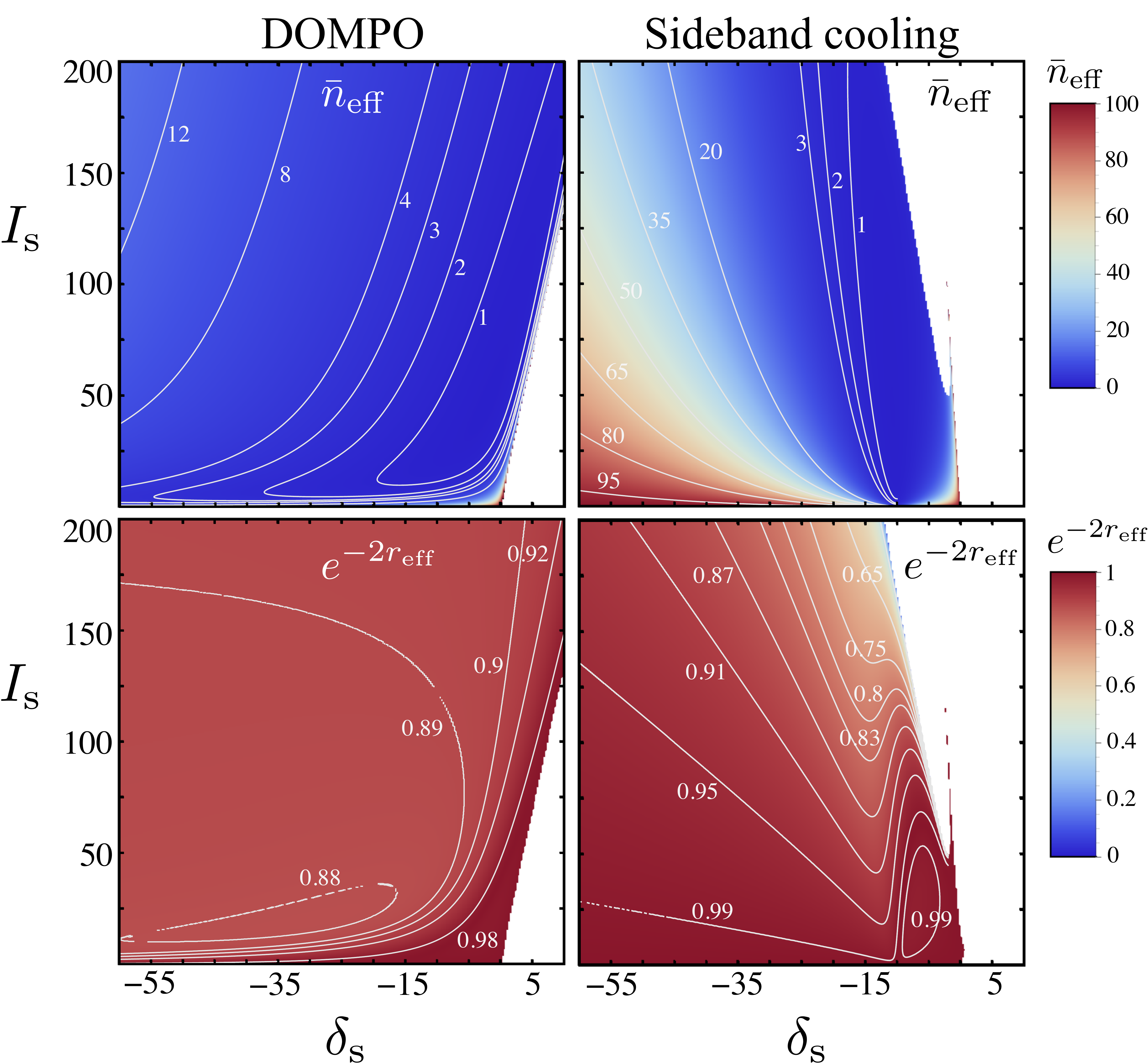}\caption{Density plots
of the effective phonon number $\bar{n}_{\mathrm{eff}}$ and mechanical
squeezing $e^{-2r_{\mathrm{eff}}}$ as a function of the signal detuning
$\delta_{\mathrm{s}}$ and intensity $I_{\mathrm{s}}$. The rest of parameters
have the same values as in Fig. \ref{FigResultsGs}, and we have fixed the
optomechanical coupling to $g=0.1$. The left panels correspond to the DOMPO,
while the right ones correspond to standard sideband cooling. It is to be
appreciated how much insensitive the cooling obtained in the DOMPO is to the
detuning, which has to be approximately fixed to $-\Omega$ (red sideband) in
a standard sideband-cooling scenario. Note that the uncoloured regions
correspond to areas of the parameter space where the classical stationary
solution is unstable, and hence linearization cannot be applied.}%
\label{FigResultsDs}%
\end{figure*}

We take as an example the parameters of Fig. \ref{Fig2}, for which we already
presented the stable and unstable regions of the classical stationary
solution. In Fig. \ref{FigResultsGs} we show the variation of the effective
thermal phonon number $\bar{n}_{\mathrm{eff}}$ and squeezing
$e^{-2r_{\mathrm{eff}}}$ with $g$, as we move above threshold, that is,
$I_{\mathrm{s}}>0$. It can be appreciated how the mechanical state is deeply
affected by the optomechanical interaction. In particular, we see that the
effective phonon number can decrease to low values, what shows the ability of
the down-converted field to cool down the mechanical motion. Moreover, as the
optomechanical coupling is enhanced, the effective squeezing levels of the thermal mechanical
state increase, up to about 50\% of squeezing in the figure. This opens the
possibility of using the DOMPO to generate squeezed mechanical states.

In Fig. \ref{FigResultsDs} we compare the cooling obtained in our system
with the one that would be obtained in a standard sideband cooling scenario. In
particular, for a fixed optomechanical coupling $g$, we show $\bar
{n}_{\mathrm{eff}}$ and $e^{-2r_{\mathrm{eff}}}$ as a function of the signal
detuning $\delta_{\mathrm{s}}$ and $I_{\mathrm{s}}$, for both the DOMPO and
standard sideband cooling. We have studied the latter case by considering the standard scenario consisting of a single driven optical mode interacting with the mechanics, described by the (normalized) quantum Langevin equations
\begin{eqnarray}\label{LangNormEqsSidebandCooling}
\frac{d\hat{x}}{d\tau}  &=& \Omega^{2}\hat{p},
\\
\frac{d\hat{p}}{d\tau}  &=& -\gamma\hat{p}-\hat{x}+2g\hat{b}_{\mathrm{s}%
}^{\dagger}\hat{b}_{\mathrm{s}}+\sqrt{\frac{4\gamma\bar{n}_{\mathrm{th}}%
}{\Omega}}g_{\mathrm{DC}}\hat{p}_{\mathrm{in}}(\tau),\nonumber
\\
\frac{d\hat{b}_{\mathrm{s}}}{d\tau}  &=& E-(1-\mathrm{i}\delta_{\mathrm{s}%
}-\mathrm{i}g\hat{x})\hat{b}_{\mathrm{s}}+\sqrt{2}g_{\mathrm{DC}}\hat{b}_{\mathrm{s}%
,\mathrm{in}}(t),\nonumber
\end{eqnarray}
where $E$ is the (normalized) amplitude associated to the laser which drives the optical mode. Note that $g_\mathrm{DC}$ appears here just because we are using the same normalization as in the rest of the paper, see Eqs. (\ref{NormVars}), which is convenient for the sake of comparison; in any case, the final results are independent of this parameter, as can be appreciated from the form of the covariance matrix in Eq. (\ref{CovMat}). Collecting the relevant operators in the vector $\mathbf{\hat{b}}=\text{col}(\hat{x},\hat{p},\hat{b}_{\mathrm{s}},\hat{b}_{\mathrm{s}}^{\dagger})$, these equations take the linearized form (\ref{LinLan}) with a noise term $\widehat{\mathbf{f}}(\tau)$ without the pump components, and with a linear stability matrix
\begin{equation}
\mathcal{L}=\left(
\begin{array}
[c]{cccc}%
0 & \Omega^{2} & 0 & 0\\
-1 & -\gamma & 2g\bar{\beta}_{\mathrm{s}}^{\ast} & 2g\bar{\beta}_{\mathrm{s}%
}\\
\mathrm{i}g\bar{\beta}_{\mathrm{s}} & 0 & -(1-\mathrm{i}\delta_{\mathrm{s}%
}-\mathrm{i}g\bar{x}) & 0\\
-\mathrm{i}g\bar{\beta}_{\mathrm{s}}^{\ast} & 0 & 0 & -(1+\mathrm{i}%
\delta_{\mathrm{s}}+\mathrm{i}g\bar{x})
\end{array}
\right),
\end{equation}
which is the usual one found in the standard optomechanical case \cite{Marquardt07,Genes08}. Following the approach explained above, we obtain the covariance matrix (\ref{CovMat}) associated to this problem, and from it, the effective thermal phonon number and effective mechanical squeezing.

The differences in the cooling performance for these two systems can be appreciated in Fig. \ref{FigResultsDs}. In particular, it is apparent that the cooling obtained in the DOMPO is less sensitive to the detuning, which in standard sideband cooling needs to be tuned to the red sideband $\delta_{\mathrm{s}}\approx-\Omega$. This insensitivity to the system parameters is something that we have also observed for other parameters, and in other regions of the parameter space, and it constitutes a main difference between the DOMPO and standard sideband optomechanical cooling.

\section{Conclusions and outlook}

In conclusion, in this work we have analyzed the DOMPO from two approximate (but relevant) perspectives: the classical limit and the linearized theory of quantum fluctuations. From a fundamental point of view, such study has been motivated by the question of how does a quantum-correlated field affect the mechanical motion. From a practical viewpoint, the study is timely because it is to be expected that the analyzed model will be experimentally implemented soon, in particular in the form of crystalline whispering gallery mode resonators or superconducting circuits.

\btxtb{We have made an exhaustive analysis of the classical phase diagram, which provides highly relevant information prior to the application of more accurate quantum techniques (see \cite{Degenfeld15b} for an example in this direction)}. Our results show that the optomechanical interaction has the effect of introducing new dynamical instabilities not present in the DOPO, as well as changing the location of the instabilities already present in it.

As for the quantum properties, when working above threshold the linearized theory has shown the ability of the quantum-correlated (squeezed) field to cool down the mechanical motion, not only to a regular thermal state, but also to a squeezed thermal state as the optomechanical coupling is enhanced. Moreover, such cooling has been shown to be more insensitive to parameters (most prominently detuning) than the one obtained through standard sideband cooling. Unfortunately, in the three-mode problem \cnb{defined by} the DOMPO it is very challenging to get analytical results from the linearization technique, \btxtb{or even to get a conclusive physical picture for the observed phenomenology}. Therefore, it will be interesting to apply other techniques which might clarify the physical processes underlying the results presented here. Techniques such as the adiabatic elimination of the optical modes in the master equation of the system might be key to this purpose.

Finally, we emphasize again on the failure of the linearization theory below threshold, where the optomechanical interaction becomes purely nonlinear. This opens an even more interesting venue, since it is a problem which will require more elaborate techniques capable of capturing nongaussian effects. Particularly relevant for us is the c-MoP theory \cite{Degenfeld14} that we recently applied to the DOPO problem \cite{Degenfeld15}, which also has the virtue of regularizing the results at the critical points of the classical theory. With the help of such an approach we showed in a recent work \cite{Degenfeld15b} that, even below threshold, the down-converted field can cool down the mechanical motion, through a process that we identified by a ``cooling by heating'' mechanism \cite{Mari12}. It will be interesting to analyze then whether this is also the mechanism responsible for the cooling that we observe above threshold in this work.

We have benefited from discussions with Yue Chang, Christoph Marquardt, Eugenio Rold\'{a}n, Tao Shi, Germ\'{a}n J. de Valc\'{a}rcel, Joaqu\'{\i}n Ruiz-Rivas, Michael J. Hartmann, Florian Marquardt, Vittorio Peano, and J. Ignacio Cirac. S.P.-O. thanks the Theory Division of the Max-Planck Institute of Quantum Optics for their support and hospitality. F.J. thanks the DFG Collaborative Research Center TRR 109, \textquotedblleft Discretization in Geometry and Dynamics\textquotedblright. P.D.-S. thanks the German Research Foundation (DFG) for support via the CRC 631 and the grant HA 5593/3-1. C.N.-B. acknowledges the financial support of the Alexander von Humboldt Foundation through its Fellowship for Postdoctoral Researchers.

\appendix

\section{Details of the stability analysis on the classical solutions}\label{StabApp}

We provide in this appendix all the details concerning our treatment of the instabilities present in the DOMPO model, which we summarized in Sec. \ref{LinStabAn}.

\subsection{Stability of the trivial solution\label{BTstab}}

In the case of the trivial stationary solution ($\bar{\beta}_{\mathrm{s}}=0$), the linear stability matrix (\ref{Lmatrix}) is highly simplified, acquiring in particular a
box structure $\mathcal{L}=\mathcal{L}_{\mathrm{m}}\oplus\mathcal{L}_{\mathrm{p}}\oplus\mathcal{L}_{\mathrm{s}}$, where the second block is already in diagonal form
\begin{equation}
\mathcal{L}_{\mathrm{p}}=\left(
\begin{array}
[c]{cc}%
-\kappa(1-\mathrm{i}\delta_{\mathrm{p}}) & 0\\
0 & -\kappa(1+\mathrm{i}\delta_{\mathrm{p}})
\end{array}
\right)  \text{,}%
\end{equation}
and its two eigenvalues have negative real part, the first block is given by%
\begin{equation}
\mathcal{L}_{\mathrm{m}}=\left(
\begin{array}
[c]{cc}%
0 & \Omega^{2}\\
-1 & -\gamma
\end{array}
\right)  ,
\end{equation}
whose eigenvalues $\lambda_{\mathrm{m}}^{(\pm)}=-(\gamma\pm\sqrt{\gamma^{2}-4\Omega^{2}})/2$ have also negative real part, and finally the last block reads
\begin{equation}
\mathcal{L}_{\mathrm{s}}=\left(
\begin{array}
[c]{cc}%
-1+\mathrm{i}\delta_{\mathrm{s}} & \bar{\beta}_{\mathrm{p}}\\
\bar{\beta}_{\mathrm{p}}^{\ast} & -1-\mathrm{i}\delta_{\mathrm{s}}%
\end{array}
\right)  ,
\end{equation}
with eigenvalues%
\begin{equation}
\lambda_{\mathrm{s}}^{(\pm)}=-1\pm\sqrt{I_{\mathrm{p}}-\delta_{\mathrm{s}}%
^{2}}.
\end{equation}
Hence, as commented in the text, we see that $\lambda_{\mathrm{s}}^{(+)}$ provides a static instability of the trivial solution, located at $I_{\mathrm{p}}=1+\delta_{\mathrm{s}}^{2}$.

\subsection{Stability of the nontrivial solution}

In the case of the nontrivial solution the $6\times6$ linear stability matrix (\ref{Lmatrix}) does not have a box structure, and hence their eigenvalues do
not have simple analytic expressions. However, we are not as interested in the actual eigenvalues as we are in the points where the real part of some of them becomes zero, since those are the points marking the instabilities, and these points can be found by analyzing the characteristic polynomial of the stability matrix, which we write as $P(\lambda)=\sum_{n=0}^{6}c_{n}\lambda^{n}$. Most of the coefficients $c_{n}(I_{\mathrm{s}},\delta_{\mathrm{s}},\delta_{\mathrm{p}},g,\kappa,\gamma,\Omega)$ are quite lengthy, and hence we don't show them here, except for the independent one, which can be written as $c_{0}=4q_{2}I_{\mathrm{s}}+2q_{1}$, where $q_{1}$ and $q_{2}$ are defined in (\ref{qcoefs}).

Given the characteristic polynomial, the static instabilities can be found from the condition $P(\lambda=0)=0$, that is, they are located in the region of the parameter space defined by the equation $c_{0}=0$, which in our case gives $I_{\mathrm{s}}=I_{\mathrm{s}}^{\mathrm{TP}}$ as mentioned in the text. We can then try to do the same with the Hopf bifurcations, but in that case the expressions are not as easy to handle. It is instructive to first consider
the case without optomechanical coupling, $g=0$. In this case the characteristic polynomial can be factorized as $P(\lambda)=P_{\mathrm{DOPO}}(\lambda)P_{\mathrm{m}}(\lambda)$, where $P_{\mathrm{m}}(\lambda)=\lambda^{2}+\gamma\lambda+\Omega^{2}$ is the characteristic polynomial associated to the free mechanical motion (hence showing no instabilities), while $P_{\mathrm{DOPO}}(\lambda)=\sum_{n=0}^{4}d_{n}\lambda^{n}$, with
\begin{eqnarray}
d_{0}  &=& \kappa^{2}I_{\mathrm{s}}(I_{\mathrm{s}}+2-2\delta_{\mathrm{p}}\delta_{\mathrm{s}}),
\\
d_{1} &=& 2\kappa\lbrack I_{\mathrm{s}}+\kappa(1+I_{\mathrm{s}}+\delta_{\mathrm{p}}^{2})],\nonumber
\\
d_{2} &=& \kappa\lbrack4+2I_{\mathrm{s}}+\kappa(1+\delta_{\mathrm{p}}^{2})],\nonumber
\\
d_{3} &=& 2(1+\kappa),\nonumber
\\
d_{4} &=& 1,\nonumber
\end{eqnarray}
is the characteristic polynomial associated to the optical modes coupled through the parametric down-conversion process, that is, to the DOPO \cite{Lugiato88,Pettiaux89,Fabre90}. The Hopf instabilities are found by locating the points in the parameter space where the eigenvalues become purely imaginary, $\lambda=\mathrm{i}\omega_{\mathrm{HB}}$, where the real parameter $\omega_{\mathrm{HB}}$ is known as the \textit{Hopf frequency} (providing the frequency of the periodic solution which is born right at the bifurcation). Applying this condition to the DOPO's characteristic polynomial, we get
\begin{eqnarray}
P_{\mathrm{DOPO}}(\lambda=\mathrm{i}\omega_{\mathrm{HB}})&=&(d_{0}-d_{2}\omega_{\mathrm{HB}}^{2}+d_{4}\omega_{\mathrm{HB}}^{4})\nonumber
\\
&&+\mathrm{i}\omega_{\mathrm{HB}}(d_{1}-d_{3}\omega_{\mathrm{HB}}^{2})=0; \label{PdopoEq}%
\end{eqnarray}
the imaginary part of this equation provides us with the Hopf frequency
\begin{equation}
\omega_{\mathrm{HB}}^{2}=\frac{d_{1}}{d_{3}}=\frac{\kappa\lbrack
I_{\mathrm{s}}+\kappa(1+I_{\mathrm{s}}+\delta_{\mathrm{p}}^{2})]}{1+\kappa},
\end{equation}
which is well defined for every value of the parameters, while the real part of (\ref{PdopoEq}) provides the condition $d_{0}d_{3}^{2}+d_{4}d_{1}^{2}%
-d_{2}d_{1}d_{3}=0$, which can be solved for $I_{\mathrm{s}}$ analytically, leading to the simple expression of Eq. (\ref{IpHB}) provided in the text.

In the $g\neq0$ case the large order of the characteristic polynomial has prevented us from finding simple analytic expressions for the dynamical instabilities of the DOMPO. Let us then pass now to explain how we have dealt with them. Proceeding as in the previous case, the real and imaginary parts of $P(\lambda=\mathrm{i}\omega_{\mathrm{HB}})$ provide us with two coupled equations
\begin{subequations}
\begin{align}
c_{0}-c_{2}\omega_{\mathrm{HB}}^{2}+c_{4}\omega_{\mathrm{HB}}^{4}-c_{6}%
\omega_{\mathrm{HB}}^{6}  &  =0,\label{HopfReal}\\
\omega_{\mathrm{HB}}(c_{1}-c_{3}\omega_{\mathrm{HB}}^{2}+c_{5}\omega
_{\mathrm{HB}}^{4})  &  =0. \label{HopfImaginary}%
\end{align}
\end{subequations}
We can see that $\omega_{\mathrm{HB}}=0$ and $c_{0}=0$ is a solution of the equations, that is, they contain the pitchfork bifurcation, what is not surprising since they are general and valid for any type of instability. Now, for $\omega_{\mathrm{HB}}\neq0$, we can proceed as follows. The second equation (\ref{HopfImaginary}) can be solved for the Hopf frequency as
\begin{equation}
\omega_{\mathrm{HB,\pm}}^{2}=\frac{c_{3}\pm\sqrt{c_{3}^{2}-4c_{1}c_{5}}%
}{2c_{5}}; \label{wHB}%
\end{equation}
these solutions can be introduced in (\ref{HopfReal}), but unfortunately the resulting equations do not allow to find a simple analytic solution for
$I_{\mathrm{s}}$. However, a symbolic program such as Mathematica allows us to find analytic solutions, provided that we write the equation as a more
manageable polynomial. In particular, let us write $\omega_{\mathrm{HB,\pm}}^{2}=l\pm r$ with $l=c_{3}/2c_{5}$ $\ $and $r=\sqrt{c_{3}^{2}-4c_{1}c_{5}}/2c_{5}$, which allows us to rewrite (\ref{HopfReal}) as
\begin{eqnarray}
c_{0}-c_{2}l+c_{4}(l^{2}+r^{2})&&-c_{6}(l^{3}+3lr^{2})
\\
&&=\pm r[c_{2}-2c_{4}l+c_{6}(3l^{2}+r^{2})].\nonumber
\end{eqnarray}
The square of this expression provides a sixth order polynomial equation for $I_{\mathrm{s}}$, whose solutions can be efficiently handled by a symbolic program. Note that by taking the square of the previous equation, we are indeed introducing extra fictitious solutions for $I_{\mathrm{s}}$, but we have checked that these extra solutions are always complex, and hence they do not provide anything which could be interpreted as instabilities. This procedure has allowed us to make an exhaustive numerical analysis of the Hopf instabilities for $g\neq0$, of which we have shown a characteristic example in Fig. \ref{Fig2}.

\section{Evaluation of the covariance matrix and the effective mechanical parameters}\label{CovApp}

In this section we explain a route to find the covariance matrix of the DOMPO directly from the linearized quantum Langevin
equations (\ref{LinLan}). We also prove expression (\ref{EffPar}) for the effective mechanical thermal phonon number and squeezing.

In order to find the covariance matrix $V$ we proceed as follows. First, we find the
left-eigenvectors of the linear stability matrix. These can be computed as the
eigenvectors of its transpose, $\{\mathcal{L}^{T}\mathbf{w}_{j}=\lambda
_{j}\mathbf{w}_{j}\}_{j=1,2,...,6}$, from which we build the matrix
$\mathcal{W}=\operatorname{col}(\mathbf{w}_{1}^{T},\mathbf{w}_{2}%
^{T},...,\mathbf{w}_{6}^{T})$, as well as the diagonal matrix of eigenvalues
$\Lambda=\mathrm{diag}(\lambda_{1},\lambda_{2},...,\lambda_{6})$. With these
definitions, we have $\mathcal{WL}=\Lambda\mathcal{W}$. Hence, applying
$\mathcal{W}$ on the left of the linearized Langevin equations (\ref{LinLan}),
and defining the vector $\mathbf{\hat{c}}(\tau)=\mathcal{W}\delta
\mathbf{\hat{b}}(\tau)$, we get a set of uncoupled linear equations for its
components, leading to the solution%
\begin{equation}
\mathbf{\hat{c}}(\tau)=\sqrt{2}g_{\mathrm{DC}}\int_{0}^{\tau}d\tau^{\prime
}e^{\Lambda(\tau-\tau^{\prime})}\mathcal{W}\widehat{\mathbf{f}}(\tau^{\prime}),
\end{equation}
in the asymptotic limit $\tau\gg\max_{j=1,2,...,6}\operatorname{Re}%
\{\lambda_{j}\}^{-1}$. It is then straightforward to compute the corresponding
correlation matrix $\mathcal{C}$, with elements $C_{jl}(\tau)=\langle\hat
{c}_{j}(\tau)\hat{c}_{l}(\tau)\rangle$, which in the asymptotic limit read%
\begin{equation}
C_{jl}=-2g_{\mathrm{DC}}^{2}\frac{\left(  \mathcal{WMW}^{T}\right)  _{jl}%
}{\lambda_{j}+\lambda_{l}},
\end{equation}
where the matrix $\mathcal{M}$ is defined in (\ref{Mmatrix}).

On the other hand, the quadrature vector $\mathbf{\hat{r}}$ is related to the
vector\ $\mathbf{\hat{b}}$ by $\mathbf{\hat{r}}=\mathcal{R}\mathbf{\hat{b}}$, where $\mathcal{R}=\mathcal{R}_{\mathrm{m}%
}\oplus\mathcal{R}_{\mathrm{p}}\oplus\mathcal{R}_{\mathrm{s}}$, with%
\begin{eqnarray}
\mathcal{R}_{\mathrm{m}}&=&\frac{1}{g_\mathrm{DC}}\left(
\begin{array}
[c]{cc}%
1/\sqrt{\Omega} & 0\\
0 & \sqrt{\Omega}%
\end{array}
\right),
\\
\mathcal{R}_{\mathrm{p}}&=&\frac{1}{g_\mathrm{DC}\sqrt{\kappa}}\left(
\begin{array}
[c]{cc}%
1 & 1\\
-\mathrm{i} & \mathrm{i}%
\end{array}
\right),\nonumber
\\
\mathcal{R}_{\mathrm{s}}&=&\frac{1}{g_\mathrm{DC}}\left(
\begin{array}
[c]{cc}%
1 & 1\\
-\mathrm{i} & \mathrm{i}%
\end{array}\right), \nonumber
\end{eqnarray}
and hence, its fluctuations\ can be written as $\delta\mathbf{\hat{r}}%
(\tau)=\mathcal{RW}^{-1}\mathbf{\hat{c}}(\tau)$, leading
to the final form of the covariance matrix in the asymptotic limit
\begin{equation}
V=\mathcal{RW}^{-1}(\mathcal{C}+\mathcal{C}^{T}%
)\mathcal{W}^{-1T}\mathcal{R}^{T}\text{.} \label{CovMat}%
\end{equation}
This is the expression that we have used to compute the Gaussian steady state of the system, which can be efficiently evaluated numerically for any value of the
parameters, since it just requires diagonalizing the $6\times6$ matrix $\mathcal{L}^{T}$.

Let us now pass to derive the relation between the effective mechanical parameters and the elements of the mechanical covariance matrix. In order to find it, we just need to realize that the thermal state is a Gaussian state with covariance matrix $V_{\mathrm{th}}(\bar{n}_{\mathrm{eff}})=(2\bar{n}_{\mathrm{eff}}+1)\mathcal{I}_{2\times2}$, where $\mathcal{I}_{2\times2}$ is the $2\times2$ identity matrix, while the squeezing operator simply acts as the symplectic transformation $\mathcal{S}(r_{\mathrm{eff}})=\mathrm{diag}(e^{-r_{\mathrm{eff}}},e^{r_{\mathrm{eff}}})$ in the space of covariance matrices \cite{Weedbrook12,NavarreteQI}. Hence, the Gaussian state corresponding to (\ref{RhoEff}) has a diagonal covariance matrix
\begin{eqnarray}\label{Vdiag1}
\tilde{V}_{\mathrm{m}}(\bar{n}_{\mathrm{eff}},r_{\mathrm{eff}})&=&\mathcal{S}%
(r_{\mathrm{eff}})V_{\mathrm{th}}(\bar{n}_{\mathrm{eff}})\mathcal{S}%
^{T}(r_{\mathrm{eff}})
\\
&=&(2\bar{n}_{\mathrm{eff}}+1)\mathrm{diag}%
(e^{-2r_{\mathrm{eff}}},e^{2r_{\mathrm{eff}}}).\nonumber
\end{eqnarray}
Therefore, the phase-shift $\hat{R}(\theta)$ applied to $\hat{\rho}_{\mathrm{m}}$ in Eq. (\ref{RhoEff})
is nothing but the rotation that diagonalizes $V_{\mathrm{m}}$, turning
it into
\begin{equation}\label{Vdiag2}
\tilde{V}_{\mathrm{m}}=\mathrm{diag}(V_{-},V_{+}),
\end{equation}
where the eigenvalues of $V_{\mathrm{m}}$ are given by $V_{\mp}=\mathrm{tr}%
\{V_{\mathrm{m}}\}/2\mp\sqrt{\mathrm{tr}\{V_{\mathrm{m}}\}^{2}/4-\det
\{V_{\mathrm{m}}\}}$. In other words, $\theta$ is the angle in phase space
that captures the direction in which squeezing is applied to the mechanical
motion. Matching the expressions (\ref{Vdiag1}) and (\ref{Vdiag2}) for the diagonal forms of $\tilde
{V}_{\mathrm{m}}$ provides the expressions (\ref{EffPar}) written in the text.


\begin{thebibliography}{99}

\bibitem {Boyd}R. W. Boyd, \textit{Nonlinear Optics }(Academic, 2003).

\bibitem {BlueBook}P. Meystre and D. F. Walls (eds.), \textit{Nonclassical
Effects in Quantum Optics} (American Institute of Physics, 1991).

\bibitem {NavarretePhDthesis}C. Navarrete-Benlloch, arXiv:1504.05917.

\bibitem {CarmichaelBook2}H. J. Carmichael, \textit{Statistical Methods in
Quantum Optics 2}, Springer Verlag, Berlin (2008).

\bibitem {Takeno07}Y. Takeno, M. Yukawa, H. Yonezawa, and A. Furusawa, Opt.
Express \textbf{15}, 4321-4327 (2007).

\bibitem {Vahlbruch08}H. Vahlbruch, M. Mehmet, S. Chelkowski, B. Hage, A.
Franzen, N. Lastzka, S. Gossler, K. Danzmann, and R. Schnabel, Phys. Rev.
Lett. \textbf{100}, 033602 (2008).

\bibitem {Eberle10}T. Eberle, S. Steinlechner, J. Bauchrowitz, V. Handchen, H.
Vahlbruch, M. Mehmet, H. Muller-Ebhardt, and R. Schnabel, Phys. Rev. Lett.
\textbf{104}, 251102 (2010).

\bibitem {Mehmet10}M. Mehmet, H. Vahlbruch, N. Lastzka, K. Danzmann, and R.
Schnabel, Phys. Rev. A \textbf{81}, 013814 (2010).

\bibitem{Ligo11} The LIGO Scientific Collaboration, Nat. Phys. \textbf{7}, 962 (2011).

\bibitem {Goda08}K. Goda, O. Miyakawa, E. E. Mikhailov, S. Saraf, R. Adhikari,
K. McKenzie, R. Ward, S. Vass, A. J. Weinstein, and N. Mavalvala, Nat. Phys.
\textbf{4}, 472-476 (2008).

\bibitem {Vahlbruch05}H. Vahlbruch, S. Chelkowski, B. Hage, A. Franzen, K.
Danzmann, and R. Schnabel, Phys. Rev. Lett. \textbf{95}, 211102 (2005).

\bibitem {Treps03}N. Treps, N. Grosse, W. P. Bowen, C. Fabre, H.-A. Bachor,
and P. K. Lam, Science \textbf{301}, 940-943 (2003).

\bibitem {Treps02}N. Treps, U. Andersen, B. Buchler, P. K. Lam, A.
Ma\^{\i}tre, H.-A. Bachor, and C. Fabre, Phys. Rev. Lett. \textbf{88}, 203601 (2002).

\bibitem {Braunstein05}S. L. Braunstein and P. van Loock, Rev. Mod. Phys.
\textbf{77}, 513-577 (2005).

\bibitem {Weedbrook12}C. Weedbrook, S. Pirandola, R. Garc\'{\i}a-Patron, N. J.
Cerf, T. C. Ralph, J. H. Shapiro, and S. Lloyd, Rev. Mod. Phys. \textbf{84},
621-669 (2012).

\bibitem {NavarreteQI} C. Navarrete-Benlloch, \textit{An Introduction to the Formalism of Quantum Information with Continuous Variables} (Morgan \& Claypool Publishers and IOP Publishing, 2015).

\bibitem {Kippenberg07}T. J. Kippenberg and K. J. Vahala, Opt. Exp.
\textbf{15}, 17172 (2007).

\bibitem {Marquardt09}F. Marquardt and S. M. Girvin, Physics \textbf{2}, 40 (2009).

\bibitem {Meystre13}P. Meystre, Annalen der Physik \textbf{525}, 215 (2013).

\bibitem {Aspelmeyer13}M. Aspelmeyer, T. J. Kippenberg, and F. Marquardt, Rev.
Mod. Phys. \textbf{86}, 1391 (2014).

\bibitem {Gigan06}S. Gigan, H. R. B\"{o}hm, M. Paternostro, F. Blaser, G.
Langer, J. B. Hertzberg, K. C. Schwab, D. B\"{a}uerle, M. Aspelmeyer, and A.
Zeilinger, Nature \textbf{444}, 67 (2006).

\bibitem {Arcizet06}O. Arcizet, P.-F. Cohadon, T. Briant, M. Pinard, and A.
Heidmann, Nature \textbf{444}, 71 (2006).

\bibitem {Schliesser06}A. Schliesser, P. Del'Haye, N. Nooshi, K. J. Vahala,
and T. J. Kippenberg, Phys. Rev. Lett. \textbf{97}, 243905 (2006).

\bibitem {Corbitt07}T. Corbitt, Y. Chen, E. Innerhofer, H. M\"{u}ller-Ebhardt,
D. Ottaway, H. Rehbein, D. Sigg, S. Whitcomb, C. Wipf, and N. Mavalvala, Phys.
Rev. Lett. \textbf{98}, 150802 (2007).

\bibitem {Thompson08}J. D. Thompson, B. M. Zwickl, A. M. Jayich, F. Marquardt,
S. M. Girvin, and J. G. E. Harris, Nature \textbf{452}, 72 (2008).

\bibitem {Wilson09}D. J. Wilson, C. A. Regal, S. B. Papp, and H. J. Kimble,
Phys. Rev. Lett. \textbf{103}, 207204 (2009).

\bibitem {Teufel11}J. D. Teufel, T. Donner, D. Li, J. W. Harlow, M. S. Allman,
K. Cicak, A. J. Sirois, J. D. Whittaker, K. W. Lehnert, and R. W. Simmonds,
Nature \textbf{475}, 359 (2011).

\bibitem {Chan11}J. Chan, T. P. Mayer Alegre, A. H. Safavi-Naeini, J. T.
Hill1, A. Krause1, S. Gr\"{o}blacher, M. Aspelmeyer, and O. Painter, Nature
\textbf{478}, 89 (2011).

\bibitem {Karuza12}M. Karuza, C. Molinelli, M. Galassi, C. Biancofiore, R.
Natali, P. Tombesi, G. Di Giuseppe, and D. Vitali, New J. Phys. \textbf{14},
095015 (2012).

\bibitem{Harris15} M. Underwood, D. Mason, D. Lee, H. Xu, L. Jiang, A. B. Shkarin, K. B\o{}rkje, S. M. Girvin, J. G. E. Harris, Phys. Rev. A \textbf{92}, 061801(R).

\bibitem {WilsonRae07}I. Wilson-Rae, N. Nooshi, W. Zwerger, and T. J.
Kippenberg, Phys. Rev. Lett. \textbf{99}, 093901 (2007).

\bibitem {Marquardt07}F. Marquardt, J. P. Chen, A. A. Clerk, and S. M. Girvin,
Phys. Rev. Lett. \textbf{99}, 093902 (2007).

\bibitem {Genes08}C. Genes, D. Vitali, P. Tombesi, S. Gigan, and M.
Aspelmeyer, Phys. Rev. A \textbf{77}, 033804 (2008).

\bibitem {Marshall03}W. Marshall, C. Simon, R. Penrose, and D. Bouwmeester,
Phys. Rev. Lett. \textbf{91}, 130401 (2003).

\bibitem {Kleckner08}D. Kleckner, I. Pikovski, E. Jeffrey, L. Ament, E. Eliel,
J. van den Brink, and D. Bouwmeester, New J. Phys. \textbf{10}, 095020 (2008).

\bibitem {Romero-Isart11}O. Romero-Isart, A. C. Pflanzer, F. Blaser, R.
Kaltenbaek, N. Kiesel, M. Aspelmeyer, and J. I. Cirac, Phys. Rev. Lett.
\textbf{107}, 020405 (2011).

\bibitem {Romero-Isart12}O. Romero-Isart, Phys. Rev. A \textbf{84}, 052121 (2012).

\bibitem {Fabre94}C. Fabre, Phys. Rev. A \textbf{49}, 1337 (1994).

\bibitem {Mancini94}S. Mancini and P. Tombesi, Phys. Rev. A \textbf{49}, 4055 (1994).

\bibitem {Brooks12}D. W. C. Brooks, T. Botter, S. Schreppler, T. P. Purdy, N.
Brahms, and D. M. Stamper-Kurn, Nature \textbf{488}, 476 (2012).

\bibitem {Safavi13}A. H. Safavi-Naeini, S. Gr\"{o}blacher, J. T. Hill, J.
Chan, M. Aspelmeyer, and O. Painter, Nature \textbf{500}, 185 (2013).

\bibitem {Weis10}S. Weis, R. Rivi\`{e}re, S. Del\'{e}glise, E. Gavartin, O.
Arcizet, A. Schliesser, and T. J. Kippenberg, Science \textbf{330}, 1520 (2010).

\bibitem {Safavi11omit}A. H. Safavi-Naeini, T. P. M. Alegre, J. Chan, M.
Eichenfield, M. Winger, Q. Lin, J. T. Hill, D. E. Chang, and O. Painter,
Nature \textbf{472}, 69 (2011).

\bibitem {Teufel11omit}J. D. Teufel, D. Li, M. S. Allman, K. Cicak, A. J.
Sirois, J. D. Whittaker, and R. W. Simmonds, Nature \textbf{471}, 204 (2011).

\bibitem {Massel12}F. Massel, S. U. Cho, J.-M. Pirkkalainen, P. J. Hakonen,
T.T.Heikkil\"{a}, and M. A. Sillanp\"{a}\"{a}, Nat. Commun. \textbf{3}, 987 (2012).

\bibitem {Karuza13}M. Karuza, C. Biancofiore, M. Bawaj, C. Molinelli, M.
Galassi, R. Natali, P. Tombesi, G. Di Giuseppe, and D. Vitali, Phys. Rev. A
\textbf{88}, 013804 (2013).

\bibitem {Rabl11}P. Rabl, Phys. Rev. Lett. \textbf{107}, 063601 (2011).

\bibitem {Stannigel10}K. Stannigel, P. Rabl, A. S. S\o rensen, P. Zoller, and
M. D. Lukin, Phys. Rev. Lett. \textbf{105}, 220501 (2010).

\bibitem {Safavi11}A. H Safavi-Naeini and O. Painter, New J. Phys.
\textbf{13}, 013017 (2011).

\bibitem {Regal11}C. A. Regal and K. W. Lehnert, J. Phys.: Conference Series
\textbf{264}, 012025 (2011).

\bibitem {Taylor11}J. M. Taylor, A. S. S\o rensen, C. M. Marcus, and E. S.
Polzik, Phys. Rev. Lett. \textbf{107}, 273601 (2011).

\bibitem {Wang12PRL}Y.-D. Wang and A. A. Clerk, Phys. Rev. Lett. \textbf{108},
153603 (2012).

\bibitem {Wang12}Y.-D. Wang and A. A. Clerk, New J. Phys. \textbf{14}, 105010 (2012).

\bibitem {Barzanjeh12}Sh. Barzanjeh, M. Abdi, G. J. Milburn, P. Tombesi, and
D. Vitali, Phys. Rev. Lett. \textbf{109}, 130503 (2012).

\bibitem {Bochmann13}J. Bochmann, A. Vainsencher, D. D. Awschalom and A. N.
Cleland, Nature Phys. \textbf{9}, 712 (2013).

\bibitem {Bagci14}T. Bagci, A. Simonsen, S. Schmid, L. G. Villanueva, E.
Zeuthen, J. Appel, J. M. Taylor, A. S\o rensen, K. Usami, A. Schliesser, and
E. S. Polzik, Nature \textbf{507}, 81 (2014).

\bibitem {Andrews14}R. W. Andrews, R. W. Peterson, T. P. Purdy, K. Cicak, R.
W. Simmonds, C. A. Regal, and K. W. Lehnert, Nature Phys. \textbf{10}, 321 (2014).

\bibitem {SchliesserPhDthesis}A. Schliesser, \textit{Cavity Optomechanics and
Optical Frequency Comb Generation with Silica Whispering-Gallery-Mode
Microresonators} (PhD thesis, Ludwig-Maximilians-Universit\"{a}t M\"{u}nchen, 2009).

\bibitem{Palomaki13} T. A. Palomaki, J. W. Harlow, J. D. Teufel, R. W. Simmonds, and K. W. Lehnert, Nature \textbf{495}, 210 (2013).

\bibitem{Agarwal09} S. Huang and G. S. Agarwal, Phys. Rev. A \textbf{79}, 013821 (2009).

\bibitem{Agarwal09bis} S. Huang and G. S. Agarwal, Phys. Rev. A \textbf{80}, 033807 (2009).

\bibitem{Marquardts15} V. Peano, H. G. L. Schwefel, Ch. Marquardt, and F. Marquardt, Phys. Rev. Lett. \textbf{115}, 243603 (2015).

\bibitem{Agarwal16} G. S. Agarwal and S. Huang, arXiv:1602.02214.

\bibitem{Nori15} X.-Y. L\"u, Y. Wu, J. R. Johansson, H. Jing, J. Zhang, and F. Nori, Phys. Rev. Lett. \textbf{114}, 093602 (2015). 

\bibitem{Xuereb12} A. Xuereb, M. Barbieri, and M. Paternostro, Phys. Rev. A \textbf{86}, 013809 (2012).

\bibitem {Ilchenko03}V. S. Ilchenko, A. A. Savchenkov, A. B. Matsko, and L.
Maleki, J. Opt. Soc. Am. B \textbf{20}, 333 (2003).

\bibitem {Ilchenko04}V. S. Ilchenko, A. A. Savchenkov, A. B. Matsko, and L.
Maleki, Phys. Rev. Lett. \textbf{92}, 043903 (2004).

\bibitem {Savchenkov07}A. A. Savchenkov, A. B. Matsko, M. Mohageg, D. V.
Strekalov, and L. Maleki, Opt. Lett. \textbf{32}, 157 (2007).

\bibitem {Furst10}J. U. F\"{u}rst, D. V. Strekalov, D. Elser, M. Lassen, U. L.
Andersen, C. Marquardt, and G. Leuchs, Phys. Rev. Lett. \textbf{104}, 153901 (2010).

\bibitem {Furst10b}J. U. F\"{u}rst, D. V. Strekalov, D. Elser, A. Aiello, U.
L. Andersen, Ch. Marquardt, and G. Leuchs, Phys. Rev. Lett. \textbf{105},
263904 (2010).

\bibitem {Hofer10}J. Hofer, A. Schliesser, and T. J. Kippenberg, Phys. Rev, A
\textbf{82}, 031804(R) (2010).

\bibitem {Furst11}J. U. F\"{u}rst, D. V. Strekalov, D. Elser, A. Aiello, U. L.
Andersen, Ch. Marquardt, and G. Leuchs, Phys. Rev. Lett. \textbf{106}, 113901 (2011).

\bibitem {Beckmann11}T. Beckmann, H. Linnenbank, H. Steigerwald, B. Sturman,
D. Haertle, K. Buse, and I. Breunig, Phys. Rev. Lett. \textbf{106}, 143903 (2011).

\bibitem {Werner12}C. S. Werner, T. Beckmann, K. Buse, and I. Breunig, Opt.
Lett. \textbf{37}, 4224 (2012).

\bibitem {Fortsch13}M. F\"{o}rtsch, J. U. F\"{u}rst, C. Wittmann, D.
Strekalov, A. Aiello, M. V. Chekhova, C. Silberhorn, G. Leuchs, and C.
Marquardt, Nature Commun. \textbf{4}, 1818 (2013).

\bibitem {Marquardt13}C. Marquardt, D. Strekalov, J. F\"{u}rst, M.
F\"{o}rtsch, and G. Leuchs, Opt. Phot. News \textbf{24}, 38 (2013).

\bibitem {Fortsch14}M. F\"{o}rtsch, G. Schunk, J. U. F\"{u}rst, D. Strekalov,
T. Gerrits, M. J. Stevens, F. Sedlmeir, H. G. L. Schwefel, S. W. Nam, G.
Leuchs, and C. Marquardt, Phys. Rev. A \textbf{91}, 023812 (2015).

\bibitem {Fortsch14b}M. F\"{o}rtsch, T. Gerrits, M. J. Stevens, D. Strekalov,
G. Schunk, J. U. F\"{u}rst, U. Vogl, F. Sedlmeir, H. G. L. Schwefel, G.
Leuchs, S. W. Nam, and C. Marquardt, J. Opt. \textbf{17}, 065501 (2015).

\bibitem{Leghtas15} Z. Leghtas, S. Touzard, I. M. Pop, A. Kou, B. Vlastakis, A. Petrenko, K. M. Sliwa, A. Narla, S. Shankar, M. J. Hatridge, M. Reagor, L. Frunzio, R. J. Schoelkopf, M. Mirrahimi, M. H. Devoret, Science \textbf{347}, 853 (2015).

\bibitem{Nation15} P. D. Nation, J. Suh, and M. P. Blencowe, arXiv:1507.00115.

\bibitem{Gigan07} S. Gigan, L. Lopez, V. Delaubert, N. Treps, C. Fabre, A. Maitre, J. Mod. Opt. \textbf{53}, 809 (2006).

\bibitem {Lugiato88}L. Lugiato, C. Oldano, C. Fabre, E. Giacobino, and R. J.
Horowicz, Il Nuovo Cimento \textbf{10}, 959 (1988).

\bibitem {Pettiaux89}N. P. Pettiaux, R.-D. Li, and P. Mandel, Optics Commun.
\textbf{72}, 256 (1989).

\bibitem {Fabre90}C. Fabre, E. Giacobino, A. Heidmann, L. Lugiato, S. Reynaud,
M. Valdacchino, and W. Kaige, Quantum Opt. \textbf{2}, 159 (1990).

\bibitem{Degenfeld15b} P. Degenfeld-Schonburg, M. Abdi, M. J. Hartmann, and C. Navarrete-Benlloch, Phys. Rev. A, in press.

\bibitem {WMbook}D.F. Walls and G. Milburn, \textit{Quantum Optics}, Springer (2007).

\bibitem {ZollerBook}C. W. Gardiner and P. Zoller, \textit{Quantum Noise}, Springer
Verlag (2004).

\bibitem {Narducci88}L. M. Narducci and N. B. Abraham, \textit{Laser Physics
and Laser Instabilities} (World Scientific, Singapore, 1988).

\bibitem {Navarrete14}C. Navarrete-Benlloch, E. Rold\'{a}n, Y. Chang, and T.
Shi, Optics Express \textbf{22}, 024010 (2014).

\bibitem {Drummond80}P. D. Drummond, K. J. McNeil, and D. F. Walls, J. Mod. Opt. \textbf{28}, 211 (1981).

\bibitem {Lugiato81}L. A. Lugiato and G. Strini, Opt. Commun. \textbf{41}, 67 (1981).

\bibitem {Collett84}M. J. Collett and C. W. Gardiner, Phys. Rev. A
\textbf{30}, 1386 (1984).

\bibitem {Kinsler93}P. Kinsler, M. Fern\'{e}e, and P. D. Drummond, Phys. Rev.
A \textbf{48}, 3310 (1993).

\bibitem {Kinsler95}P. Kinsler and P. D. Drummond, Phys. Rev. A \textbf{52},
783 (1995).

\bibitem {Drummond02}P. D. Drummond, K. Dechoum, and S. Chaturvedi, Phys. Rev.
A \textbf{65}, 033806 (2002).

\bibitem {Chaturvedi99}S. Chaturvedi and P. D. Drummond, Eur. Phys. J. B
\textbf{8}, 251 (1999).

\bibitem {Pope00}D. T. Pope, P. D. Drummond, and S. Chaturvedi, Phys. Rev. A
\textbf{62}, 042108 (2000).

\bibitem {MertensPRL93}C. J. Mertens, T. A. B. Kennedy, and S. Swain Phys.
Rev. Lett., \textbf{71}, 2014 (1993).

\bibitem {MertensPRA93}C. J. Mertens, T. A. B. Kennedy, and S. Swain Phys.
Rev. A, \textbf{48}, 2374 (1993).

\bibitem {Veits97}O. Veits and M. Fleischhauer, Phys. Rev. A \textbf{55}, 3059 (1997).

\bibitem {Degenfeld15}P. Degenfeld-Schonburg, C. Navarrete-Benlloch, and M. J.
Hartmann, Phys. Rev. A \textbf{91}, 053850 (2015).

\bibitem {PositiveP}P. D. Drummond and C. W. Gardiner, J. Phys. A: Math. Gen.
\textbf{13}, 2353 (1980).

\bibitem{Degenfeld14} P. Degenfeld-Schonburg and M. J. Hartmann, Phys. Rev. B \textbf{89}, 245108 (2014).

\bibitem{Mari12} A. Mari and J. Eisert, Phys. Rev. Lett. \textbf{108}, 120602 (2012).

\end{thebibliography}
\end{document}